\documentclass[12pt, draftcls, peerreview, onecolumn]{IEEEtran}

\def\BibTeX{{\rm B\kern-.05em{\sc i\kern-.025em b}\kern-.08em
    T\kern-.1667em\lower.7ex\hbox{E}\kern-.125emX}}

\usepackage{bbm}
\usepackage{amsmath}
\usepackage{amsthm}
\usepackage{amsfonts}
\usepackage{amssymb}
\usepackage{amscd}
\usepackage{graphicx}
\usepackage{newlfont}
\usepackage{cite}
\usepackage{color}
\usepackage{multirow,tabularx}
\usepackage{ifthen}
\usepackage{enumitem}
\usepackage{times}
\usepackage{stackrel}
\usepackage{floatrow}
\usepackage{algorithm,algpseudocode}
\usepackage{epstopdf}
\usepackage[all]{xy}
\usepackage{url}
\usepackage[font = footnotesize]{caption}
\usepackage{booktabs}
\usepackage{multirow,tabularx}
\usepackage{hyphenat}
\usepackage{verbatim}
\usepackage{mathtools}
\usepackage{multicol,lipsum}
\usepackage[table]{xcolor} 
\usepackage{mathtools}
\usepackage{flexisym}
\usepackage{graphicx}
\usepackage{subfig}
\usepackage{etoolbox}

\def\BibTeX{{\rm B\kern-.05em{\sc i\kern-.025em b}\kern-.08em
        T\kern-.1667em\lower.7ex\hbox{E}\kern-.125emX}}

\IEEEoverridecommandlockouts

\DeclarePairedDelimiter\abs{\lvert}{\rvert}%

\newcommand{\norm}[1]{\left\Vert#1\right\Vert}

\newcommand{\tr}{\mathrm{tr}}

\newcommand{\Real}{\mathbb{R}}

\newcommand{\vech}{\mathbf{h}}

\newcommand{\vecx}{\mathbf{x}}

\newcommand{\vecy}{\mathbf{y}}

\newcommand{\vecw}{\mathbf{w}}

\newcommand{\matW}{\mathbf{W}}

\newcommand{\matH}{\mathbf{H}}

\newcommand{\trieq}{\triangleq}
\allowdisplaybreaks

\let\oldabs\abs
\def\abs{\@ifstar{\oldabs}{\oldabs*}}

\newcommand{\nuserpercell}{{N}} 
 
\newcommand{\ntxants}{{M}}
\newcommand{\totpow}{P_T}
\newcommand{\pow}{P}

\makeatletter
\def\endthebibliography{%
	\def\@noitemerr{\@latex@warning{Empty `thebibliography' environment}}%
	\endlist
}
\makeatother

\begin{document}
\allowdisplaybreaks

\title{A Joint Solution for Scheduling and Precoding in Multiuser MISO Downlink Channels }
 \author{$\text{Ashok Bandi}^{}$, $\text{Bhavani Shankar Mysore R, }^{}$ $\text{Symeon Chatzinotas}^{}$ $\text{and Bj{\"o}rn Ottersten}^{}$\\        
        Interdisciplinary Centre for Security, Reliability and Trust (SnT), the University of Luxembourg, Luxembourg.
        \\ Email: \{ashok.bandi, bhavani.shankar, symeon.chatzinotas, bjorn.ottersten\}@uni.lu
        \thanks{This work has been published in part at Global Conference on Signal and Information Processing 2018 \cite{GlobalSIP_ref}
    }
        \thanks{This work is supported in part by Luxembourg national fund FNR project PROSAT.
    }}


	\maketitle
	\vspace{-0.9cm}
 \begin{abstract}          
        The long-term average performance of the MISO downlink channel, with a large number of users compared to transmit antennas of the BS, depends on the interference management which necessitates the joint design problem of scheduling and precoding. Unlike the previous works which do not offer a truly joint design, this paper focuses on formulating a problem amenable for the joint update of scheduling and precoding. Novel optimization formulations are investigated to reveal the hidden difference of convex/ concave structure for three classical criteria (weighted sum rate, max-min SINR, and power minimization) and associated constraints are considered. Thereafter, we propose a convex-concave procedure framework based iterative algorithm where scheduling and precoding variables are updated jointly in each iteration. Finally, we show the superiority in performance of joint solution over the state-of-the-art designs through Monte-Carlo simulations.

    \end{abstract}
    \vspace{-0.5cm}
    \begin{keywords}  User scheduling, Precoding, Multiuser 
    \end{keywords}
\vspace{-0.6cm}
\section{Introduction}
    With the adoption of full frequency reuse in the next generation cellular networks, interference among the simultaneously served users becomes a limiting factor thwarting the achievement of near-optimal capacity\cite{1683918, Beamforming,406634,69993}. Moreover, in a network with a large number of users compared to the number of BS transmit antennas, scheduling the users for simultaneous transmission is pivotal for interference management \cite{GreedyUS_Dimic, ChanOrthoZFBF_goldsmith}. In this work, we address the joint design of scheduling and precoding problem for multiuser MISO downlink channels in single cell scenario for the following design criteria: 1) Maximize the WSR subject to user SINR, scheduling and power constraints which is referred to simply as WSR.
   2) Maximize the minimum SINR of the scheduled users subject to scheduling and total power constraints which is referred to as MMSINR. 3)  Minimize the power utilized subject to scheduling and minimum SINR constraints which is referred to as PMIN.
    
    The joint design of scheduling and precoding, which we simply refer to as joint design, is well studied for the last decade (see \cite{ResrcAllcReview_2017} and references therein). Most of the existing literature on the joint design can be classified as:
    \begin{itemize}
        \item \textit{Non-iterative decoupled approach}: In this approach, scheduling and precoding are treated as two decoupled problems where usually the users are scheduled according to some criteria followed by precoding \cite{GreedyUS_Dimic, ChanOrthoZFBF_goldsmith},\cite{SUS_limitedfeedback,8239622, MMSINR_ref}.
        \item  \textit{Iterative decoupled approach}: In this approach, scheduling and precoding are still treated as two separate problems. However, scheduling and precoding parameters are refined in each iterate to improve the objective based on the feedback from the previous iterate \cite{WYU_Multicell_JointSchPrec, MINLP_ref1, MulticellSUS,4641963}.
        \item  \textit{Joint formulation with alternate update}: In this approach, the joint design problem is formulated as a function of both scheduling and precoding \cite{SparsePrecPenalty_JointSchBF_JSAC,6334281,CS_PC_CRN_16}. However, these formulations are not amenable for the joint update and during the solution stage either scheduling constraints are ignored \cite{SparsePrecPenalty_JointSchBF_JSAC} or the scheduling and precoding variables are updated alternatively \cite{6334281}.
    \end{itemize}
    
    The joint design is a coupled problem where the efficiency of the precoder design depends on interference which, in turn, is a function of the scheduled users \cite{ResrcAllcReview_2017}. Hence, the joint update of scheduling and precoding has the potential to achieve better performance over the aforementioned approaches \cite{SparsePrecPenalty_JointSchBF_JSAC,GreedyUS_Dimic,ChanOrthoZFBF_goldsmith},\cite{SUS_limitedfeedback,8239622,MMSINR_ref,WYU_Multicell_JointSchPrec,MINLP_ref1,MulticellSUS,4641963}. The joint design problem is combinatorial and NP-hard due to scheduling; it is also non-convex due to the constraints on the SINR or rate of scheduled users \cite{WYU_Multicell_JointSchPrec}. This design further spans Boolean space (user scheduling) and continuous space (precoding vector). To alleviate the complexity of an exhaustive search for practical system dimensions motivates a shift towards low-complexity achievable solutions. In this context, we quickly review the various relevant works to place ours in perspective.
    
    The joint design problem to maximize the weighted sum rate subject to total power constraint, which is referred to as the classical WSR problem, is considered for single cell networks in \cite{GreedyUS_Dimic, ChanOrthoZFBF_goldsmith, SUS_limitedfeedback}. The channel orthogonality based scheduling followed by zero-forcing precoding (SUS-ZF) proposed in \cite{ChanOrthoZFBF_goldsmith} is proven to be asymptotically optimal for sum rate maximization. However, it is easy to see that SUS-ZF is not optimal for WSR with non-uniform weights and QoS constraints. Similarly, the classical WSR is addressed for multicell networks in \cite{WYU_Multicell_JointSchPrec,MulticellSUS,4641963} and hierarchical networks in \cite{SparsePrecPenalty_JointSchBF_JSAC}. The joint design problem is also considered for MMSINR in \cite{MMSINR_ref} and PMIN in \cite{MINLP_ref1}. However, scheduling and precoding are not jointly updated in the aforementioned works. Moreover, the coupled nature of binary variables with precoding vector arises in many other formulations \cite{6920005, SparsePrecodingPenalty_MTao} etc. For example, in \cite{SparsePrecPenalty_JointSchBF_JSAC} towards maximizing the weighted sum-rate in a hierarchical network, binary variables associated with users get multiplied to signal power and interference power of SINR. Similarly, in \cite{CS_PC_CRN_16} binary variable is multiplied to the rate of the users in weighted sum-rate maximization problem. Please note that system models and objectives discussed in \cite{CS_PC_CRN_16, SparsePrecPenalty_JointSchBF_JSAC, SparsePrecodingPenalty_MTao} are different from each other, and the emphasis is only on the occurrence of the joint design (coupled discrete and continuous) nature that prevails in different designs. The multiplicative nature in previous formulations precludes the joint update of scheduling and precoding. To the best of our knowledge, no prior work exists that update the scheduling and precoding jointly for the aforementioned WSR, MMSINR and PMIN problems. Therefore, we focus on formulating the joint design problem for WSR, MMSINR, and PMIN that facilitates the joint scheduling and precoding solutions.
    
    The WSR and MMSINR design problems for fixed scheduled users are non-convex with difficulty to obtain a global solution. However, efficient suboptimal solutions have been proposed for WSR in \cite{CC_2017} and MMSINR in \cite{6155555,6476606} by formulating these as DC programming problems with the help of auxiliary variables and SDP transformations. However, the semidefinite relaxations for WSR and MMSINR often lead to non-unity rank solutions from which the approximate rank-1 solutions are extracted \cite{CC_2017,6155555,6476606}. The rank-1 approximation results in a loss of performance. Moreover, the transformed problems have higher complexity than the original problems due to auxiliary variables and SDP transformations. In this work, we propose WSR and MMSINR problems for the joint design as DC programming problems without SDP transformation and with a minimal number of auxiliary variables.

    The aforementioned discussion reflects on the novelties of the paper-based both on problem formulation and its solution. The contributions of the paper include:    \begin{itemize}
        \item  The scheduling is handled through the power of the precoding vector of the corresponding user, where non-zero power indicates the user being scheduled and not scheduled otherwise. Unlike the previous works \cite{CS_PC_CRN_16,SparsePrecPenalty_JointSchBF_JSAC,SparsePrecodingPenalty_MTao}, a binary variable is used for upper bounding the power of the precoding vector. This renders the formulation amenable to the joint design of scheduling and precoding.
        \item With the help of the aforementioned scheduling,  the joint design problem for WSR, MMSINR, and PMIN design criteria are formulated as MINLP in a way that would facilitate the joint updates of scheduling and precoding. Here, the nonconvexity of the problem stems from rate and SINRs in the objective and constraints.
        \item The binary nature of the problem due to scheduling constraints is addressed by relaxing the binary variables into real values. This is followed by penalizing the objective with a novel entropy-based penalty function to promote a binary solution for the scheduling variables. This step transforms the optimization into a continuous non-convex problem.
        \item Unlike the classical difference-of-convex/concave (DC) formulation using SDP transformation \cite{CC_2017,6155555,6476606}, a novel useful reformulation of the objective and/or SINR constraints are proposed to manipulate the joint design as DC programming without SDP transformation. 
       \item Further, a convex-concave procedure (CCP) based low-complexity iterative algorithm is proposed for all the WSR, MMSINR and PMIN DC problems. A procedure is proposed to find the feasible initial point, which is sufficient for these algorithms to converge.
       \item Subsequently, per iteration complexity of the CCP based algorithms, is discussed. Finally, the efficiency of the proposed DC reformulations is compared to the decoupled solutions using the Monte-Carlo simulations.
    \end{itemize}
    
    The rest of the paper is organized as follows. Section~\ref{sec:system_model} presents the system model and problem formulation of WSR, MMSINR and PMIN problem. The reformulations and algorithm are proposed for WSR in Section~\ref{sec:WSR}, MMSINR in Section~\ref{sec:MMSINR} and PMIN in Section~\ref{sec:JSPPMIN}. Section~\ref{sec:simulation} presents simulation results, followed by conclusions in Section~\ref{sec:conclusions}.
    
    \emph{Notation:} Lower or upper case letters represent scalars, lower case boldface letters represent vectors, and upper case boldface letters represent matrices. $\|\cdot\|$ represents the Euclidean norm, $|\cdot|$ represents the cardinality of a set or the magnitude of a scalar, $(\cdot)^H$ represents Hermitian transpose, $\binom{a}{b}$ represents  $a$ choose $b$, $\tr\lbrace \rbrace$ represents trace and $\Real\lbrace \rbrace$ represents real operation, and s.t. is referred to as subject to and $\nabla$ represents the gradient.
    \vspace{-0.5cm}
    \section{System Model}\label{sec:system_model}
    Consider the downlink transmission of a single cell MISO system with $\nuserpercell$ users in a cell and a BS with $\ntxants$ antennas. Let $\mathbf{h}_{i} \in \mathbb{C}^{\ntxants \times 1}$, $\mathbf{w}_{i} \in \mathbb{C}^{\ntxants \times 1}$ and $x_i$ denote the downlink channel, precoding vector and data of user $i$ respectively. Let $n_i$ be the noise at user $i$. The noise at all users is assumed to be independent and characterized as  additive white complex Gaussian with zero mean and variance $\sigma^2$. Let $y_i$ be the noisy linear measurement of the user $i$ and $\vecy \trieq \left[ y_1,  \hdots y_{\nuserpercell}\right]^T$. The generative model of the measurements $\vecy$ of all users is given by
\begin{equation} \label{eq:sysmod}
\vecy = \matH\matW \vecx+ \mathbf{n},  
\end{equation}
where $\matH \trieq \left[    \mathbf{h}_1,\hdots, \vech_{\nuserpercell} \right]^H$,  $\matW \trieq \left[    \mathbf{w}_1,\hdots, \vecw_{\nuserpercell} \right]$, $\vecx \trieq \left[ x_1,  \hdots x_{\nuserpercell}\right]^T$, $\mathbf{n} \trieq \left[ n_1,  \hdots n_{\nuserpercell}\right]^T$.
   
    BS is assumed to transmit independent data to utmost $\ntxants$ among $\nuserpercell$ users and $|x_i|^2=1,\text{}\forall i$. Hence, this leads to scheduling of utmost (exactly) $\ntxants$ users for WSR (MMSINR and PMIN). 
    
    Towards defining the WSR problem mathematically, let $\mathcal{T}=\lbrace 1, \hdots \nuserpercell \rbrace$ be the set containing indices of all users and $\bar{\mathcal{K}}$ be a subset of $\mathcal{T}$ with cardinality less than or equal to $\ntxants$. Clearly, the number of possible subsets of type $\bar{\mathcal{K}}$  is $C \trieq \sum_{i=0}^{\ntxants} {\nuserpercell \choose i}$ and let $\mathcal{K}$ be the collection of all the possible subsets of type $\bar{\mathcal{K}}$. With the notations defined, the joint design problem with the objective of maximizing the WSR subject to constraint on the minimum SINR of the scheduled users and total consumed power is defined as, 
    \vspace{-0.4cm}
   \begin{align}\label{eq:WSR_def} 
    \cal{P}_{\text{WSR}}: \text{ }
    \max_{{\forall \bar{\mathcal{K}}} \in \mathcal{K}} \hspace{0.3cm}&\max_{\mathbf{W}_{\bar{\mathcal{K}}}} \sum_{\forall i\in \bar{\mathcal{K}}} \alpha_{i}R_i   \\
     &\text{s.t }  R_i \geq \epsilon_i, \forall i \in\bar{\mathcal{K}}, \nonumber  \\  &\sum_{\forall i \in \bar{\mathcal{K}}}\norm{\mathbf{w}_i}_2^2 \leq \totpow, \nonumber
    \end{align}
    $\hspace{6.5cm}\underbrace{\hspace{0.8cm}\underbrace{\hspace{3.0cm}}_{\text{precoding problem for selected users}}\hspace{0.1cm}}_{\text{Joint schedueling and Precoding problem}} $ \\
where $\alpha_{i}  \in \mathcal{R}^{+},$  $\gamma_i \triangleq \dfrac{\lvert \mathbf{h}^H_{i} \mathbf{w}_{i} \rvert^2}{\sigma^2+\sum_{j \neq i \in \bar{\mathcal{K}} }\lvert\mathbf{h}^H_{i}\vecw_{j}\rvert^2}$ and $R_i$ are weight, SINR and rate of the user $i$ respectively and $\bar{\mathcal{K}}$ is set of scheduled users, $\totpow$ is the total available power, and $\mathbf{W}_{\bar{\mathcal{K}}} = \lbrace \mathbf{w}_i \rbrace_{i \in \bar{\mathcal{K}}}^{|\bar{\mathcal{K}|}}$ is the precoding matrix containing the precoding vectors of users belonging to set $\bar{\mathcal{K}}$.

    Similarly, building towards the mathematical definition of the MMSINR - unlike the WSR design - scheduling of exactly $\ntxants$ users is considered since constraining scheduling to utmost $\ntxants$ users always leads to the trivial solution of scheduling one user.  An elaborate discussion is provided at the beginning of Section~\ref{sec:MMSINR}. Let $\bar{\mathcal{S}}$ be a subset of ${\mathcal{T}}$ with cardinality equal to $\ntxants$. Clearly, the number of possible subsets of type $\bar{\mathcal{S}}$  is ${\nuserpercell \choose \ntxants}$ and let $\mathcal{S}$ be the collection of all the possible subsets of type $\bar{\mathcal{S}}$. The design problem for MMSINR is defined as,
    \vspace{-0.3cm}
    \begin{align}\label{eq:MMSINR_def}
    \cal{P}_{\text{MMSINR}}: \text{ }
    \max_{\bar{\mathcal{S}} \subseteq \mathcal{S}}  \hspace{0.6cm}&\max_{\mathbf{W}_{\bar{\mathcal{S}}}}\hspace{0.6cm}\min_{i\subseteq \bar{\mathcal{S}}} \text{ }\lbrace \beta_i \gamma_{i} \rbrace  \\ 
 \text{     }&\text{s.t} \sum_{i \in \bar{\mathcal{S}}}\norm{\mathbf{w}_i}_2^2 \leq \totpow, \nonumber\\
 &\gamma_i \geq \epsilon_i, i \in \bar{\mathcal{S}}, \nonumber
  \hspace{-4cm}\underbrace{\hspace{0.8cm}\underbrace{\hspace{3.0cm}}_{\text{precoding problem for selected users}}\hspace{0.1cm}}_{\text{Joint schedueling and Precoding problem}} \nonumber
    \end{align}
    where $\beta_i \in \mathcal{R}^{+},$ is weight and $\mathbf{W}_{\bar{\mathcal{S}}} = \lbrace \mathbf{w}_i \rbrace_{i \in \bar{\mathcal{S}}}^{|\bar{\mathcal{S}|}}$ is the matrix containing the precoding vectors of users in the set $\bar{\mathcal{S}}$.
   
    Finally, towards defining the PMIN problem, for the same reason mentioned in MMSINR, the constraint of scheduling exactly $\ntxants$ users is considered. With notations defined for {MMSINR} criteria, and letting $\epsilon_i$ (different than $\epsilon_i$ in WSR definition) to the minimum SINR requirement of user $i$, $\forall i$, the PMIN problem is defined as: \vspace{-0.3cm}
\begin{equation}\label{eq:PMIN_def}
\cal{P}_{\text{PMIN}}: \text{ }\underbrace{ \min_{\bar{\mathcal{S}} \subseteq \mathcal{S}} \underbrace{\min_{\mathbf{W}_{\bar{\mathcal{S}}} }  \text{ } \sum_{i \in \bar{\mathcal{S}}}  \norm{\mathbf{w}_i}_2^2 \nonumber  {\text{ s.t. } \gamma_i \geq \epsilon_i,\text{ }i\subseteq {\bar{\mathcal{S}}}}.}_{\text{PMIN problem for selected users}} \nonumber 
    \vspace{-0.75cm} }_{\text{Joint user scheduling and PMIN problem}}
\end{equation}

   Notice that to accommodate the fairness in the designs, weights or priority factors are introduced through $\alpha$ and $\beta$ in WSR and MMSINR problems respectively. Various fairness metrics are proposed in the literature, e.g. fairness in terms of rates and allocated power are considered at the physical layer. We refer to \cite{Fairness_lit} and references therein for details on fairness. 
   
   The inner optimization in \eqref{eq:WSR_def}, \eqref{eq:MMSINR_def}, and \eqref{eq:PMIN_def} solves the precoding problem for the users of the selected subset. The outer optimization, on the other hand, takes care of scheduling the set with a maximum objective value among all subsets. Notice that the inner and outer optimization are coupled - the design of precoder depends on the selected set of users, while the scheduling of users depends on the objectives in \eqref{eq:WSR_def}, \eqref{eq:MMSINR_def} and \eqref{eq:PMIN_def}  which in-turn are a function of precoder \cite{MGMC_JSP}. 
   
   Towards proposing low-complexity algorithms, we begin by addressing the user scheduling through the precoding vectors. Accordingly, user $i$ is not scheduled if the norm of the corresponding precoding vector is zero i.e,
   \vspace{-0.3cm}
\begin{equation}\label{eq:SchDef}
\norm{\mathbf{w}_{i}}_2= \left\{ \begin{array}{ll}
=0; \text{user not selected,} \\
\neq 0; \text{user selected}.\\
\end{array}
\right.
\end{equation}
   The zero norm of the precoding vector $\vecw_i$ of user $i$ indicates the all elements of $\vecw_i$ are zero. Hence, the user $i$ is not scheduled. Similarly, the non-zero norm of the precoder vector $\vecw_i$ of the user $i$ indicates the user $i$ being scheduled and $\norm{\mathbf{w}_{i}}_2^2$ indicates power assigned to the user. Now, in the sequel, we focus on the design of low-complexity solutions to the joint design using \eqref{eq:SchDef} to achieve better performance than the decoupled designs. \vspace{-0.3cm}
    \section{Weighted Sum Rate maximization}\label{sec:WSR}
  In \eqref{eq:WSR_def}, the weighted sum rate objective is considered to improve the overall throughput of the network as opposed to favoring the individual users. Thus, WSR problem schedules only the users who contribute to maximizing the objective.  Given enough resources, the WSR design schedules users close to $\ntxants$ users as the weighted sum of the rates contributes linearly to the objective as opposed to the scheduling of few users with higher SINRs who contribute logarithmically to the objective. Hence, the constraint of scheduling utmost of $\ntxants$ users - unlike MMSINR and PMIN- is considered as opposed scheduling to exactly $\ntxants$ users. Besides, the design is flexible to favor users by increasing the corresponding weights i.e., $\alpha_i$ to relatively larger values over the users. The minimum rate constraints preclude scheduling of the users whose rates are not in the range of interest. Since scheduling of zero users in also included in the feasible set, the problem \eqref{eq:WSR_def} is always feasible.  In the sequel, the WSR problem (i.e., \eqref{eq:WSR_def})
   is transformed as a DC programming problem through a sequence of novel reformulations and low-complexity sub-optimal algorithms within the framework of CCP.   

   \vspace{-0.49cm}
    \subsection{Joint Design Problem Formulation: WSR}\label{sec:JointProb}
Letting $\bar{\mathcal{K}}$ to be the set of scheduled users, a tractable formulation of \eqref{eq:WSR_def} using \eqref{eq:SchDef} is,\vspace{-0.05cm}
       \begin{align}\label{eq:WSR_prob}
\cal{P}_1^{\text{WSR}}:&\text{ } \max_{\mathbf{W},\forall \bar{\mathcal{K}} \in \mathcal{K}} \hspace{0.2cm} \sum_{i=1}^{\nuserpercell}\alpha_i R_{i} \\ \text{s.t. }
C_1:& \text{    }\norm{\left[\norm{\mathbf{w}_{1}}_2,\hdots,\norm{\mathbf{w}_{\nuserpercell}}_2\right]}_0 \leq \ntxants, \text{ }  \nonumber \\            
C_2:&\text{    } \sum_{i=1}^{\nuserpercell} \norm{\mathbf{w}_{i}}_2^2 \leq \totpow,\text{ } \nonumber \\
C_3:& \text{ } R_i \geq \epsilon_i, \text{ }  i \in \bar{\mathcal{K}}. \nonumber
\end{align}  
\vspace{-0.2cm}
\emph{Remarks}:
     \begin{itemize}
     \item It is clear from \eqref{eq:SchDef} and the definition of $\ell_0$ norm, that the constraint $C_1$ imposes strict restrictions on the total number of selected users to utmost $\ntxants$. We refer to this constraint as the user scheduling constraint throughout this section.
       \item The constraint $C_2$ precludes the design from using the transmission power   greater than $\totpow.$
       \item The constraint $C_3$ imposes the minimum rate required for the scheduled users.
    \end{itemize}
   

    \emph{A Novel MINLP formulation}: The problem $\cal{P}_1^{\text{WSR}}$ is combinatorial due to the constraint $C_1$ and $C_3$, and non-convex due to the objective and constraints $C_1$ and $C_3$. Towards addressing the combinatorial nature, letting $\eta_{i}$ to be the binary scheduling variable associate with user $i$, $\boldsymbol{\eta}=[\eta_{1},\hdots,\eta_{\nuserpercell}]^T$ and $\tilde{\epsilon}_i \triangleq 2^{\epsilon_i}-1,\text{} \forall i$  a tractable formulation of $C_1$ and $C_3$ of $\cal{P}_1^{\text{WSR}}$ is,
    \vspace{-0.20cm}
    \begin{align}\label{eq:Bin_WSR_prob}
    \cal{P}_2^{\text{WSR}}:& \text{ } \max_{\mathbf{W}, \boldsymbol{\eta}} \text{ } \sum_{i=1}^{\nuserpercell}  \alpha_{i} \log\left( 1+ \gamma_i \right)
    \\ 
    \text{s.t. }  C_{1}:& \text{ }\eta_{i} \in \lbrace 0,1 \rbrace, \text{ } \forall \text{} i, \nonumber \\
    C_{2}:& \text{ } \norm{\mathbf{w}_{i}}_2^2 \text{ }\leq \text{ }{\totpow}\eta_{i}, \text{ } \forall \text{} i, \nonumber \\        
    C_{3}:& \text{ }\sum_{i=1}^{\nuserpercell} \eta_{i} \leq \ntxants,  \nonumber \\
    C_4:&  \text{ } \sum_{i=1}^{\nuserpercell}\norm{\mathbf{w}_{i}}_2^2 \leq \totpow, \nonumber \\
    C_5:& \text{ } \gamma_i \geq \eta_i \tilde{\epsilon}_i, \text{ }\forall i.\nonumber
    \end{align} 
   \vspace{-0.2cm}
    \emph{Remarks:}
    \begin{itemize}
        \item The binary nature of $\eta_{i}$ (i.e., $C_1$) together with $C_2$ determines the scheduling of users. In other words, $\eta_{i}=0$ leads to a precoding vector containing all zero entries. Similarly $\eta_{i} = 1$ leads to $\norm{\mathbf{w}_{i}}_2^2 \leq \totpow$ which is a trivial upper bound compared to $C_4$. Hence the constraint $C_2$ along with $C_1$ contributes only to the scheduling aspects of the problem.
        
        \item Constraint $C_5$ ensures minimum rate or SINR requirements of the scheduled users. If user $i$ is scheduled i.e., $\eta_i=1$, from $C_5$, $\gamma_i \geq \tilde{\epsilon}_i$. Similarly, for an unscheduled user $i$, $C_5$ becomes $\gamma_i \geq 0$. In fact for $\eta_i=1$, constraint is met with equality i.e., $\gamma_i=0$ due to $C_2$.
    \end{itemize}
    
     \emph{Novelty of $\cal{P}_2^{\text{WSR}}$}: Novelty of $\cal{P}_2^{\text{WSR}}$ lies in the formulation of scheduling constraint, $C_2$. This reformulation is vital to the facilitation of the joint update of $\boldsymbol{\eta}$ and $\mathbf{W}$ as discussed in the sequel. Kindly refer to that this formulation differs from those in the literature (\cite{SparsePrecodingPenalty_MTao,6059438,7012104, SparsePrecPenalty_JointSchBF_JSAC, CS_PC_CRN_16} etc) where the scheduling constraint is handled by a binary slack variable which multiplies either the precoding vector or the rate of the user, to control the user scheduling. This multiplication not only makes the constraints non-convex but also makes it difficult to obtain the joint update of Boolean and continuous variables due to the coupling of variables. 
    
    
     The problem $\cal{P}_2^{\text{WSR}}$ is  non-convex with combinatorial constraints where the non-convexity is due to the objective and $\boldsymbol{\eta}$, and combinatorial nature is due to $\boldsymbol{\eta}.$  Towards addressing the non-convexity, letting $\zeta_i$ to be the slack variable associated with user $i$ and $\boldsymbol{\zeta}=[\zeta_1,\hdots,\zeta_{\nuserpercell}]^T$, the problem $\cal{P}_2^{\text{WSR}}$ is equivalently reformulated as,
     \vspace{-0.4cm}
      \begin{align}
    \cal{P}_3^{\text{WSR}}:&\text{}\max_{\mathbf{W},\boldsymbol{\zeta},\boldsymbol{\eta}} \hspace{0.05cm}  f\left( \boldsymbol{\zeta},\boldsymbol{\eta}\right) \triangleq \sum_{i=1}^{\nuserpercell}  \alpha_{i} \log\left( \zeta_i \right) \label{eq:WSR_Int_DC_norm}  \\
    \text{s.t.  } C_{1}, &C_{2},  C_{3}, C_{4} \text{ in } \eqref{eq:Bin_WSR_prob}\nonumber \\
    {C}_5: & \text{ } 1+\gamma_i \geq \zeta_i, \forall i, \nonumber \\
    {C}_6: &\text{ }\zeta_i \geq 1+\eta_i \tilde{\epsilon}_i, \text{ } \forall i, \nonumber
    \end{align}
    \vspace{-0.2cm}
    \emph{Remarks:}
\begin{itemize}
\item From the objective and constraint $C_5$, the variable $\zeta_i$ provides a lower bound for $1+\gamma_i$. 
\item The constraint $C_6$ ensures minimum SINR or rate constraint of the scheduled users.
\item It is easy to see that, at the optimal solution, the constraints $C_5$ and $C_6$ are met with equality.
\end{itemize}    
 
\emph{Novelty of $\cal{P}_3^{\text{WSR}}$}: The novelty of $\cal{P}_3^{\text{WSR}}$ lies in the constraint $C_5$ which helps to reformulate the objective as a concave function and connects the minimum rate constraints to the objective. This reformulation is crucial as it facilitates the reformulation of $\cal{P}_3^{\text{WSR}}$ as DC programming problem without resorting to SDP transformations \cite{7581107,5594709,WSR_CCP_damped,CC_2017}. 
\vspace{-0.5cm}
\subsection{A Novel DC reformulation: WSR}
  A novel rearrangement of SINR constraint $C_5$ in $\cal{P}_3^{\text{WSR}}$ that  transforms $\cal{P}_3^{\text{WSR}}$ as a DC programming problem without SDP transformation is,
  \vspace{-0.3cm}
  \begin{align}
    \cal{P}_4^{\text{WSR}}:&\text{}\max_{\mathbf{W},\boldsymbol{\zeta},\boldsymbol{\eta}} \hspace{0.05cm} f\left( \boldsymbol{\zeta},\boldsymbol{\eta}\right) \triangleq \sum_{i=1}^{\nuserpercell}  \alpha_{i} \log\left( \zeta_i \right) \label{eq:DC_rearg}  \\
    \text{s.t. } C_{1}, &C_{2},  C_{3}, C_{4} \text{ and } {C}_6 \text{ in } \eqref{eq:WSR_Int_DC_norm}\nonumber \\
    {C}_5:& \text{ }  \mathcal{I}_i\left(\mathbf{W}\right) -\mathcal{G}_i\left(\mathbf{W},\zeta_i\right) \leq 0, \forall i, \nonumber
    \end{align}
 where $\mathcal{I}_i\left(\mathbf{W}\right)=\sigma^2+\sum_{j \neq i}\lvert\mathbf{h}^H_{i}\vecw_{j}\rvert^2$ and $\mathcal{G}_i\left(\mathbf{W},\zeta_i\right)=\dfrac{\sigma^2+\sum_{j = 1}^{\nuserpercell}\lvert\mathbf{h}^H_{i}\vecw_{j}\rvert^2}{\zeta_i}$. Notice that $\mathcal{I}_i\left(\mathbf{W}\right)$ is convex in $\mathbf{W}$, and for $\zeta_i >0$, $\mathcal{G}_i\left(\mathbf{W},\zeta_i\right)$ is also jointly convex in $\mathbf{W}$ and $\zeta_i$. Hence, \eqref{eq:DC_rearg} is a DC programming problem with combinatorial constraint $C_1$. This is the first attempt at reformulating the novel WSR towards a tractable form without resorting to SDP methods or additional slack variables thereby rendering the problem efficiently. 

\emph{Beyond SDP based DC formulation}:
Notice that for fixed $\boldsymbol{\eta}$, the problem $\cal{P}_3^{\text{WSR}}$ becomes a classical WSR maximization problem subject to SINR and total power constraints \cite{7581107,CC_2017,5594709,WSR_CCP_damped}. The problem $\cal{P}_3^{\text{WSR}}$ is non-convex due to the constraint $C_5$. Although, for fixed $\boldsymbol{\zeta}$, the constraint $C_5$ in $\cal{P}_3^{\text{WSR}}$ is formulated as a second-order cone programming (SOCP) constraint \cite{SOCP_Amiweisel}, the SOCP transformation of $C_5$ for a general case is not known. On the other hand, many previous works have exploited the DC structure in WSR maximization problem without SINR constraint in \cite{7581107,5594709, WSR_CCP_damped} and with SINR constraint in \cite{CC_2017} by transforming it into an SDP problem. However, the SDP transformations in \cite{7581107,5594709,WSR_CCP_damped,CC_2017}, essentially increase the number of variables hence the complexity. Moreover, SDP transformations also introduce the non-convex rank-1 constraint on the solutions which is difficult to handle in general which led to semidefinite relaxations  \cite{Bengtsson454064} followed by extraction of approximate feasible rank-1 solutions.

    The problem $\cal{P}_4^{\text{WSR}}$ is still an MINLP with the structure in the non-convexity being DC which can be leveraged with the optimization tools like CCP. Now, to circumvent the combinatorial nature of $\cal{P}_4^{\text{WSR}}$, ${\eta_{i}}$ is relaxed to a box constraint between 0 and 1, and penalized with $\mathbb{P}\left(\eta_{i}\right)$ so that the relaxed problem favours 0 or 1. The penalized reformulation of $\cal{P}_4^{\text{WSR}}$ with penalty parameter  $\lambda_1 \in \mathcal{R}^{+}$ is, \vspace{-0.4cm}
    \begin{align}\label{eq:Penalized_WSR_bin} 
    \cal{P}_5^{\text{WSR}}:&\text{ }\max_{\mathbf{W},\boldsymbol{\eta}, \boldsymbol{\zeta}} \hspace{0.1cm} \sum_{i=1}^{\nuserpercell} \left( \alpha_{i}  \log\left(1+\gamma_i\right)+ \lambda_1 \mathbb{P}\left(\eta_{i}\right) \right)
    \\ 
    \text{s.t. }{C}_1:&\text{    }  0 \leq \eta_{i} \leq 1, \text{ }\forall i, \nonumber \\
    C_2,&C_3, C_4, C_5\text{ and } C_6 \text{ in } \eqref{eq:DC_rearg}. \nonumber
    \end{align}
    We propose a new penalty function
$\mathbb{P}{\left(\eta_{i}\right)} \triangleq \eta_{i}\log\eta_{i}+\left(1-\eta_{i}\right)\log\left(1-\eta_{i}\right)$ which is a convex function in $\eta_{i} \geq 0$. $\mathbb{P}{\left(\eta_{i}\right)}$ incurs no penalty at $\eta_{i}= 0 \text{ or }1$ and the penalty increases logarithemically as $\eta_i$ drifts away from $\eta_{i}= 0 \text{ or }1$ with the highest penalty at $\eta_{i}=0.5$. Hence, by choosing $\lambda_1$ appropriately, binary nature of $\boldsymbol{\eta}$ is ensured. 

   Now, notice that the objective in $\cal{P}_5^{\text{WSR}}$ a difference of concave functions i.e. $f\left( \boldsymbol{\zeta},\boldsymbol{\eta}\right)=\sum_{i=1}^{\nuserpercell} \left( \alpha_{i} \log\left( \zeta_i \right)\right)- \left(-\sum_{i=1}^{\nuserpercell} \lambda \mathbb{P}\left(\eta_{i}\right) \right)$ and constraints are convex and DC. Hence, the problem $\cal{P}_5^{\text{WSR}}$ is a DC programming problem. In the sequel, a CCP based algorithm is proposed\cite{Yuille_CCCP_2001}.
   \vspace{-0.5cm}
   \subsection{JSP-WSR: A Joint Design Algorithm}\label{sec:wsr_Prop_algo}
    In this section, we propose a CCP  based iterative algorithm to the DC problem in \eqref{eq:Penalized_WSR_bin} which we refer to as \emph{JSP-WSR}. CCP is a powerful tool to find a stationary point of DC programming problems. Within this framework, an iterative procedure is performed, wherein the two steps of Convexification and Optimization are executed in each iteration. In the convexification step, a concave optimization problem is obtained from $\cal{P}_5^{\text{WSR}}$ by linearizing the objective and constraints. Hence, by definition, the concavified objective and convexified constraints lower bound the objective and constraints of $\cal{P}_5^{\text{WSR}}$ where the lower bound is tight at the previous iteration. The optimization step then solves the convex subproblem globally. Thus, the proposed JSP-WSR algorithm iteratively executes the following two steps until convergence:
    \begin{itemize}
        \item Convexification: Let ${\mathbf{W}}^{k-1}, \boldsymbol{\eta}^{k-1}, \boldsymbol{\zeta}^{k-1}$ be the estimates of $\mathbf{W}, \boldsymbol{\eta}, \boldsymbol{\zeta}$ in iteration $k-1$ and $\mathcal{G}_i(\mathbf{W},\zeta^{}_{i})$. In iteration $k$, the convex part of the objective in $\cal{P}_5^{\text{WSR}}$, $\left(-\sum_{i=1}^{\nuserpercell} \lambda \mathbb{P}\left(\eta_{i}\right) \right)$, and the concave part of constraint $C_5$ in $\cal{P}_{11}^{\text{MM}}$ are replaced by their first order Taylor approximations around the estimate of  $\left({\mathbf{W}}^{k-1}, \boldsymbol{\eta}^{k-1}, \boldsymbol{\zeta}^{k-1}\right)$ 
        \begin{align}\label{eq:WSR_convx}
        &\tilde{\mathbb{P}}\left(\eta_i\right) \triangleq \nonumber \text{ }\lambda \left(\mathbb{P}\left(\eta^{k-1}_{i}\right) + \left(\eta_{i}-\eta_{i}^{k-1}\right) \nabla \mathbb{P}\left(\eta_{i}^{k-1}\right) \right), \nonumber\\
        &\tilde{\mathcal{G}}_i({\mathbf{W}^{k-1}},\zeta^{k-1}_{i}) \triangleq  -\mathcal{G}_i(\mathbf{W},\zeta_{i})- \Real \left\{ \tr \left\{\nabla^H \mathcal{G}_i(\mathbf{W}^{k-1},\zeta^{k-1}_{i}) \begin{bmatrix}
    {\mathbf{w}_1-\mathbf{w}_1^{k-1}} \\
    \vdots \\
    {\mathbf{w}_{\nuserpercell}-\mathbf{w}_{\nuserpercell}^{k-1}} \\
    {\zeta{i}-\zeta^{k-1}_{i}}
    \end{bmatrix} \right\}  \right\},
    \end{align}
    where  
    \begin{equation}\label{eq:Grad}
    \nabla \mathcal{G}_i(\mathbf{W}^{k-1},\zeta^{k-1}_{i}) = \begin{bmatrix}
    \dfrac{2{\mathbf{h}_i\mathbf{h}_i^{H}\mathbf{w}_1^{k-1}}}{\zeta_i^{k-1}} \\
    \vdots \\
    \dfrac{2{\mathbf{h}_{i}\mathbf{h}_{i}^{H}\mathbf{w}_{\nuserpercell}^{k-1}}}{\zeta_i^{k-1}} \\
    -\dfrac{\sigma^2+\sum_{j = 1}^{\nuserpercell}\lvert\mathbf{h}^H_{i}\vecw_{j}^{k-1}\rvert^2}{{\zeta_i^{k-1}}^2}
    \end{bmatrix}.
    \end{equation}
        
        \item Optimization: The next update $\left({\mathbf{W}}^{k+1}, \boldsymbol{\eta}^{k+1},\boldsymbol{\zeta}^{k+1}\right)$ is obtained by solving the following convex problem (which is obtained by replacing convex part of the objective and constraints in $\cal{P}_5$ with \eqref{eq:WSR_convx} and ignoring the constant terms in the objective) : 
        \vspace{-0.1cm}
        \begin{align}\label{eq:WSR_DC_sol}
        \cal{P}_6^{\text{WSR}}:&\text{}\max_{\mathbf{W},\boldsymbol{\zeta},\boldsymbol{\eta}} \hspace{0.1cm}  \sum_{i=1}^{\nuserpercell} \left( \alpha_{i} \log\left( \zeta_i \right)+ \lambda_1  \eta_i \nabla \mathbb{P}\left(\eta_{i}^{k-1}\right) \right)  \\ 
        \text{s.t } C_{1},& C_{2}, C_{3}, C_{4}\text{ and } C_{6} \text{ in } \eqref{eq:Penalized_WSR_bin}\nonumber \\
        {C}_5:&\text{ } \mathcal{I}_i\left(\mathbf{W}\right) -\tilde{\mathcal{G}}_i({\mathbf{W}},\zeta_{i})\leq 0, \forall i. \nonumber
        \end{align}
    \end{itemize}
    
    JSP-WSR is a CCP based iterative algorithm; hence, the complexity of the algorithm depends on complexity of the sub-problems $\cal{P}_6^{\text{WSR}}$. The convex problem $\cal{P}_6^{\text{WSR}}$ has $\left(\nuserpercell \ntxants+2\nuserpercell \right)$ decision variables and $\left(2\nuserpercell+1 \right)$ convex constraints and $2\nuserpercell+1$ linear constraints. Hence, the computational complexity of $\cal{P}_6^{\text{WSR}}$ is $\mathcal{O}\left(\left(\nuserpercell \ntxants+2\nuserpercell \right)^3 \left(4\nuserpercell+2 \right) \right)$ \cite{Complexity_ref}
    
    Note that the proposed JSP-WSR algorithm is based on CCP framework hence a feasible initial point (FIP) is sufficient for the CCP procedure to converge to a stationary point (kindly refer \cite{Sriperumbudur_CCP_2009}). In many cases, obtaining a FIP is difficult. However, in the next section, we propose a method which promises to obtain at least one FIP.
    \vspace{-0.5cm}
    \subsection{Feasible Initial Point: {WSR}}\label{sec:FIP_WSR}
     CCP is an iterative algorithm and an initial feasible point guarantees the solutions of all iterations remain feasible. In many cases, it is difficult find a feasible initial Let $\mathbf{1}$ and  $\mathbf{0}$ be column vectors of length $\nuserpercell$ with all ones and zeros respectively. A trivial initial FIP is obtained by the initializing $\lbrace\mathbf{w}_i=\boldsymbol{0}\rbrace_{i=1}^{\nuserpercell}, \boldsymbol{\eta}=\boldsymbol{0}$ and $\boldsymbol{\zeta}=\boldsymbol{1}$. Perhaps, a better FIP could be obtained by the following iterative procedure.
     \begin{itemize}
     \item Step 1: Initialize $\boldsymbol{\eta}=\boldsymbol{\hat{\eta}}$ that satisfies constraints $C_1$ and $C_3$ in $\cal{P}_5$, $\boldsymbol{\zeta}=\boldsymbol{1}$ and $0<\delta<1$.
     \item Step 2:  Solve the following optimization: 
    \vspace{-0.2cm}\begin{align}\label{eq:FIP_WSR} 
    \cal{P}_{\text{FESWSR}}: &\text{ }\lbrace \hat{\mathbf{W}}\rbrace:\text{ } \text{find} \text{ }{\mathbf{W}} \hspace{0.05cm} 
   \\
   \text{s.t. } 
 \tilde{C}_{1}:& \hspace{0.05cm} \norm{\mathbf{w}_{i}}_2^2 \text{ }\leq \text{ }\hat{\eta_{i}}{\totpow}, \text{ } \forall \text{} i, \nonumber \\      
   \tilde{C}_2:& \hspace{0.05cm} \norm{\left[\sigma \hdots \lbrace\mathbf{h}^H_{i}\vecw_{j}\rbrace_{j\neq i} \hdots\right]}_2 \leq \dfrac{\mathbf{h}^H_{i}\vecw_{i}}{\sqrt{ \hat{\eta_i} \tilde{\epsilon}_i} }, \text{ }\forall i,\nonumber\\
    \tilde{C}_3:& \hspace{0.05cm} \Real\lbrace\mathbf{h}^H_{i}\vecw_{i} \rbrace \geq 0, \forall i, \nonumber \\
   \tilde{C}_4:& \hspace{0.05cm} \Im\lbrace\mathbf{h}^H_{i}\vecw_{i} \rbrace == 0, \forall i, \nonumber \\
 \tilde{C}_5:& \hspace{0.05cm} \norm{\mathbf{W}}_2^2 \leq \totpow. \nonumber
   \end{align}
     \item Step 3: If $\hat{\mathbf{W}}$ is feasible go to step 4 else update $\boldsymbol{{\eta}}=\delta \boldsymbol{\hat{\eta}}$ and go to step 2.
     \item Step 4: Choose $\hat{\zeta}_i$ such that $1+\hat{\eta}_i\tilde{\epsilon}_i \leq \hat{\zeta}_i \leq  1+\hat{\gamma}_i$ where $\hat{\gamma}_i$ is the SINR of the user $i$ calculated using $\Hat{\mathbf{W}}$.
   \end{itemize}
   \vspace{-0.15cm}
   \emph{Remarks:}
   \begin{itemize}
   \item Notice that the updates of $\hat{\boldsymbol{\eta}}$ are always feasible. Different $\hat{\boldsymbol{\eta}}$ in step 1 which satisfy the constraint $C_1$ and $C_3$ in $\cal{P}_5$ may lead to different FIPs. Similarly, different choices of $\delta \in \left(0,1\right)$ in step 1 may also lead to different FIPs.
   \item The optimization problem in Step 2 is only a function of $\mathbf{W}$ since $\boldsymbol{\eta}$ is fixed apriori and $\boldsymbol{\zeta}$ can be calculated easily from the solution given in step 4.
   \item This method always gives an initial feasible point since updates of $\boldsymbol{\eta}$ eventually lead to $\boldsymbol{\hat{\eta}}=\mathbf{0}$ and thus $\Hat{\mathbf{W}}$ in step 2 becomes feasible with $\hat{\mathbf{W}}=0$. By initializing $\hat{\boldsymbol{\eta}}$ close to $\mathbf{0}$, FIP can be obtained in fewer iterations. 
   \item The FIP obtained by this procedure may not be feasible for the original WSR problem $ \cal{P}_{\text{WSR}}$ in \eqref{eq:WSR_def} unless $\Hat{\mathbf{W}}$ becomes feasible for $\lbrace \hat{\eta_i}\in \lbrace0,1\rbrace\rbrace_{i=1}^{\nuserpercell}$ satisfying $\sum_{i=1}^{\nuserpercell}\hat{\eta}_i\leq \ntxants$.
   \item Although the FIP obtained by this method is not feasible for $\cal{P}_{\text{WSR}}$, the final solution obtained by JSP-WSR with this FIP becomes a feasible for $\cal{P}_{\text{WSR}}$ since the solution satisfies the scheduling and SINR constraints of $\cal{P}_{\text{WSR}}$.
   \end{itemize}
   Letting $\cal{P}_6^{\text{WSR}}\left(k\right)$ be the objective value of the problem $\cal{P}_6^{\text{WSR}}$ at  iteration $k$, the pseudo code of JSP-WSR for the joint design problem is given in algorithm~\ref{alg:JSP_WSR}.
   \vspace{-0.3cm}
\begin{algorithm}
 \caption{JSP-WSR}
 \label{alg:JSP_WSR}
 \begin{algorithmic}[]
 \State{\textbf{Input}: $\mathbf{H},\left[\epsilon_1,\hdots,\epsilon_{\nuserpercell}\right],\totpow,\Delta$, $\boldsymbol{\eta}^0, {\mathbf{W}}^0, \lambda_1=0, k=1$}
 \State{\textbf{Output}: $\mathbf{W},\boldsymbol{\eta}$}
 \While{$|\cal{P}_6^{\text{WSR}}\left(k\right)-\cal{P}_6^{\text{WSR}}\left(k-1\right)|\geq \Delta$}
    \State \textbf{Convexification:} Convexify the problem \eqref{eq:WSR_convx}
    \State \textbf{Optimization}: Update $\left({\mathbf{W}}^{k}, \boldsymbol{\eta}^{k}, \boldsymbol{\zeta}^K\right)$ by solving $\cal{P}_5^{\text{WSR}}$ 
    \State \textbf{Update :} $\cal{P}_6\left(k\right), \lambda_1, k$
 \EndWhile
\end{algorithmic}
\end{algorithm}  
\vspace{-0.1cm}
    
\section{Max Min SINR }\label{sec:MMSINR}
 In this section, we focus on the development of a low-complexity algorithm for the MMSINR problem defined in \eqref{eq:MMSINR_def}. Dropping a user with low SINR improves minimum SINR (MSINR) as it reduces the interference to the other users and the power of the dropped user can be used to further improve the MSINR of other users. Hence, the constraint of scheduling utmost $\ntxants$ users leads to the global solution which has highest MSINR which is achieved by scheduling only one user. To avoid this, scheduling exactly $\ntxants$ users is considered for MMSINR design. Besides the scheduling constraint, the minimum SINR requirements of the scheduled users are also considered. Without the minimum SINR requirement, the design becomes superficial as the solution might include zero SINR or SINR  values which are not usable in practice. However, problem $\eqref{eq:MMSINR_def}$ may not be feasible for an arbitrary $\totpow$ for a given $\ntxants$ and $\nuserpercell$ due to the constraint of scheduling exactly $\ntxants$ users and the minimum SINR constraint on scheduled users.
 Hence, it is assumed that problem $\cal{P}_{\text{MMSINR}}$ has at least one feasible solution. A low-complexity sub-optimal algorithm using the frame work of CCP is developed for the MMSINR problem in the sequel.
 \vspace{-0.5cm}
 \subsection{Joint Design Problem Formulation: MMSINR}
 A tractable mathematical formulation of \eqref{eq:MMSINR_def} is,  
\begin{align}\label{eq:MaxMin_prob}
\cal{P}_1^{\text{MM}}:&\text{ } \max_{\mathbf{W}} \hspace{0.2cm} \min_{i=\lbrace 1, \hdots, \nuserpercell \rbrace}  \lbrace \beta_i \gamma_{i} \rbrace \\ \text{s.t. }
C_1:& \text{    }\norm{\left[\norm{\mathbf{w}_{1}}_2,\hdots,\norm{\mathbf{w}_{\nuserpercell}}_2\right]}_0 == \ntxants, \text{ }  \nonumber \\            
C_2:&\text{    } \sum_{i=1}^{\nuserpercell} \norm{\mathbf{w}_{i}}_2^2 \leq \totpow,\text{ } \nonumber \\
C_3:&\text{ } \beta_i \gamma_i \geq \mathbbm{1}\left( \norm{\mathbf{w}_i}_2\right) \epsilon_i, \nonumber
\end{align}    
where $\mathbbm{1}\left( \norm{\mathbf{w}_i}_2\right)=0$ if $\norm{\mathbf{w}_i}_2 =0$ otherwise $\mathbbm{1}\left( \norm{\mathbf{w}_i}_2\right)=1$.

 The SINR $\gamma_i$ is non-convex and piece-wise minimum of $\lbrace  \gamma_i \rbrace_{i=1}^{\nuserpercell}$ is also non-convex. So, $\cal{P}_1^{\text{MM}}$ maximizes a non-convex objective subject to a combinatorial constraint $C_1$, which is generally an NP-hard problem. Moreover obtaining a global solution to $\cal{P}_1^{\text{MM}}$ requires an exhaustive search over all the possible sets and solving the classical MMSINR problem for each set. 

\emph{Adopting classical epigraph formulation}: In the classical MMSINR problem, for the predefined selected users, SINRs of all users is addressed with a slack variable, say $s$, that lower bounds $\beta_i\gamma_i, \text{ }\forall i$ i.e., $\lbrace \beta_i\gamma_i \rbrace_{i=1}^{\nuserpercell} \geq s$\cite{5762643,7583660}. However, this approach can not be applied to a joint design problem because there are always $\nuserpercell-\ntxants$ users who are not scheduled hence their SINR must be equal to zero. Therefore, lower bounding all $\lbrace \beta_i \gamma_i \rbrace_{i=1}^{\nuserpercell}$ with $s$, makes the problem trivial and the solution, say $s^{*}$, is always zero. Letting $s$ to be a slack variable and $\mathcal{S}$ to be the set of scheduled users, adopting the epigraph formulation the problem $\cal{P}_1^{\text{MM}}$ is reformulated as, \vspace{-0.35cm}
\begin{align}\label{eq:MaxMin_Epi}
\cal{P}_2^{\text{MM}}:&\text{ } \max_{\mathbf{W}, s,\mathcal{S}} \hspace{0.2cm} s\\ \text{s.t.  } C_1,&\text{ } C_2 \text{ in } \eqref{eq:MaxMin_prob} \nonumber \\
 C_3:& \text{ }  \beta_i\gamma_i \geq s, \text{ }\forall i \in \mathcal{S}, \nonumber \\
 C_4:& \text{ } s \geq \epsilon_i , \text{ }\forall i \in \mathcal{S}, \nonumber
\end{align}

\emph{A Novel Reformulation}: Similar to WSR problem, letting $\eta_i$ to be a binary variable associated to user $i$, an equivalent formulation of $\cal{P}_2^{\text{MM}}$, without the set notation is,
\vspace{-0.3cm}
\begin{align}\label{eq:MaxMin_bin}
\cal{P}_3^{\text{MM}}:&\text{ } \max_{\mathbf{W},\boldsymbol{\eta},s} \hspace{0.2cm} s \\ \text{s.t. } 
C_1:& \text{    }\eta_i \in \lbrace 0, 1\rbrace, \text{ } \forall i, \nonumber \\
C_2:& \text{    }\norm{\mathbf{w}_{i}}_2^2 \leq \eta_i \totpow, \text{ }  \nonumber \\
C_3:& \text{    } \sum_{i=1}^{\nuserpercell} \eta_i == \ntxants, \text{ }  \nonumber \\
C_4:& \text{    } \sum_{i=1}^{\nuserpercell} \norm{\mathbf{w}_{i}}_2^2 \leq \totpow,\text{ } \nonumber \\
C_5:& \text{ }\beta_i \gamma_i \geq \eta_i \epsilon_i, \text{}\forall i, \nonumber \\
C_6:& \text{ }\beta_i \gamma_i \geq \eta_i s, \text{}\forall i. \nonumber 
\end{align}   
\vspace{-0.2cm}
\emph{Remarks}:
  \begin{itemize}
        \item Constraint $C_5$ is the minimum SINR constraint equivalently written with the help of $\eta_i$s. 
        \item The variable $s$ in $C_6$ is active only when $\eta_i=1$. For example, when user $i$ not scheduled i.e., $\eta_i=0$, its SINR is lower bounded by 0 which is satisfied always by the definition of SINR. Similarly, when user $i$ scheduled i.e., $\eta_i=1$, its SINR is lower bounded by $s$. Hence maximizing $s$ maximizes only the minimum SINR of scheduled users. 
    \end{itemize}
    \vspace{-0.5cm}
\subsection{A Novel DC reformulation: MMSINR}
The problem $\cal{P}_3^{\text{MM}}$ is a MINLP where the non-convexity is due to constraints $C_5$ and $C_6$, and combinatorial nature is due to constraint $C_1$  hence the aforementioned comments still valid. similar to constraint $C_5$ of $\cal{P}_4^{\text{WSR}}$, constraint $C_5$ of the problem $\cal{P}_3^{\text{MM}}$ can be formulated as a DC constraint. However, the same approach can not be applicable to constraint $C_6$ in $\cal{P}_3^{\text{MM}}$ as $\eta_i$ and $s$ are both variables. Moreover, to the best of our knowledge DC reformulation of constraints of type $C_6$ in $\cal{P}_3^{\text{MM}}$ is not known. In this section, a novel procedure is proposed to transform constraints of type $C_6$ in $\cal{P}_3^{\text{MM}}$ as DC constraints. This procedure involves the change of variable $s$ by $\dfrac{1}{t}$ followed by rearrangement as given below,
\begin{align}
& \beta_i \gamma_i \geq \dfrac{\eta_i}{t} \implies 1+ \beta_i \gamma_i \geq 1+\dfrac{\eta_i}{t}
\Rightarrow \mathcal{L}_i\left(\mathbf{W},t \right)-\mathcal{H}_i\left(\mathbf{W}, \eta_i, t \right) \leq 0, \label{eq: DC_constr}
\end{align}
where $\mathcal{L}_i\left(\mathbf{W},t \right)=\dfrac{\mathcal{I}_i\left(\mathbf{W}\right)}{t}$ and $\mathcal{H}_i\left(\mathbf{W}, \eta_i, t \right)=\dfrac{\mathcal{I}_i\left(\mathbf{W}\right)+ \beta_i \lvert \mathbf{h}^H_{i} \mathbf{w}_{i} \rvert^2}{t+\eta_i}$. Notice that, given $t>0$, $\mathcal{L}_i\left(\mathbf{W},t \right)$ is jointly convex in $\mathbf{W}$ and $t$ and $\mathcal{H}_i\left(\mathbf{W}, \eta_i, t \right)$ is also jointly convex in $\mathbf{W},\eta_i$ and $t$. Hence, \eqref{eq: DC_constr} is a DC constraint.

 Letting $\mathcal{J}_i\left(\mathbf{W}, \eta_i, t \right)=\dfrac{\mathcal{I}_i\left(\mathbf{W}\right)+ \beta_i \lvert \mathbf{h}^H_{i} \mathbf{w}_{i} \rvert^2}{1+\eta_i\epsilon_i}$, for the sake of completion, with the help of variable $t$ and \eqref{eq: DC_constr}, the problem $\cal{P}_3^{\text{MM}}$ is reformulated as, \vspace{-0.35cm}
\begin{align}\label{eq:MaxMin_DC_bin_Epi}
\cal{P}_{4}^{\text{MM}}:&\text{ } \min_{\mathbf{W},\boldsymbol{\eta}, t} \hspace{0.2cm} {t}  \\ \text{s.t. } C_1,& C_2, C_3, C_4\text{ in } \eqref{eq:MaxMin_bin}, \nonumber \\
C_5:& \text{ }\mathcal{I}_i\left(\mathbf{W}\right)-\mathcal{J}_i\left(\mathbf{W}, \eta_i, t \right) \leq 0, \forall i, \nonumber \\
C_6:& \text{ } \mathcal{L}_i\left(\mathbf{W},t \right)-\mathcal{H}_i\left(\mathbf{W}, \eta_i, t \right) \leq 0, \text{            } \forall i,  \nonumber \\
C_7:&\text{ } t>0. \nonumber \nonumber
\end{align}

The problem $\cal{P}_{4}^{\text{MM}}$ is a DC problem with combinatorial constraint $C_1$. To circumvent the combinatorial nature, following the approach in \ref{sec:WSR}, the binary constraint ${\eta_{i}}$ is relaxed to a box constraint between 0 and 1 and ${\eta_{i}}$ is penalized with $\mathbb{P}\left(\eta_{i}\right)$ as, \vspace{-0.3cm}
\begin{align}\label{eq:Penalized_MMSINR_bin} 
\cal{P}_{5}^{\text{MM}}:&\text{ } \min_{\mathbf{W},\boldsymbol{\eta}, t} \hspace{0.2cm} {t} - \lambda_2 \mathbb{P} \left( \eta_{i} \right)   \\ \text{s.t. } C_1:& \text{ }0 \leq \eta_i \leq 1, \text{} \forall i, \nonumber \\
C_2,& C_3, C_4, C_5, C_6, C_7 \text{ in } \eqref{eq:MaxMin_DC_bin_Epi}, \nonumber
\end{align}
where $\lambda_2 \in \mathcal{R}^{+}$ is a penalty parameter of the design. 



The problem $\cal{P}_{5}^{\text{MM}}$ maximizes a convex objective subject to convex and DC constraints. Hence $\cal{P}_{5}^{\text{MM}}$ is a DC problem and a CCP based algorithm could be solved with an FIP obtained from \ref{sec:FIP_MM} . However, the strict equality constraint $C_3$ in $\cal{P}_{5}^{\text{MM}},$  limits the update of the $\boldsymbol{\eta}$. In order to allow the flexibility in choosing $\boldsymbol{\eta}$, the following problem is considered instead:\vspace{-0.15cm}
\begin{align}\label{eq:Pen_relx_MMSINR_bin} 
\cal{P}_{6}^{\text{MM}}:\text{ } \min_{\mathbf{W},\boldsymbol{\eta}, t} \hspace{0.2cm} {t} - \lambda_2 \mathbb{P} \left( \eta_{i} \right) + \Omega \left( \sum_{i=1}^{\nuserpercell} \eta_i-\ntxants \right)^2 \text{s.t. } 
C_1, C_2, C_4, C_5, C_6, C_7 \text{ in } \eqref{eq:Penalized_MMSINR_bin}, 
\end{align}    
where $\Omega  \in \mathcal{R}^{+}$ is a penalty parameter. It is easy to see that choosing the appropriate $\Omega$ (usually higher value) ensures the equality constraint. The problem ${\cal{P}_{6}}^{\text{MM}}$ is also a DC problem and a CCP based algorithm, JSP-MMSINR, is proposed in the sequel to solve it efficiently.

\vspace{-0.5cm}
\subsection{JSP-MMSINR: A Joint Design Algorithm}\label{sec:Prop_algo_MMSINR}
In this section, we propose a CCP framework based iterative algorithm to the problem $\cal{P}_{6}^{\text{MM}}$, which is referred to as JSP-MMSINR, wherein the JSP-MMSINR executes the following Convexification and Optimization steps in each iteration:
\begin{itemize}
    \item Convexification: Let $\left({\mathbf{W}}, \boldsymbol{\eta}, t\right)^{k-1}$ be the estimates of $\left( \mathbf{W}_{i}, \eta_{i},t\right)$ in iteration $k-1$. In iteration $k$, the concave part of $C_5$ and $C_6$ in $\cal{P}_{6}^{\text{MM}}$ i.e., $-\mathcal{H}_i(\mathbf{W},\eta_{i}, t)$ and $-\mathcal{J}_i(\mathbf{W},\eta_{i}, t)$ are replaced by its affine approximation around $\left({\mathbf{W}}, \boldsymbol{\eta}, t\right)^{k-1}$ which is given by, \vspace{-0.3cm}
    \begin{align}\label{eq:MMSINR_convx}
        &\tilde{\mathcal{H}}_i\left({\mathbf{W}}, \boldsymbol{\eta}, t\right)^{k-1} \triangleq -\mathcal{H}_i\left({\mathbf{W}}, \boldsymbol{\eta}, t\right)^{k-1}- \Real \left\{ \tr \left\{\nabla^H \mathcal{H}_i\left({\mathbf{W}}, \boldsymbol{\eta}, t\right)^{k-1} \begin{bmatrix}
    {\mathbf{w}_1-\mathbf{w}_1^{k-1}} \\
    \vdots \\
    {\mathbf{w}_{\nuserpercell}-\mathbf{w}_{\nuserpercell}^{k-1}} \\
    {\eta_{i}-\eta^{k-1}_{i}} \\
    t-t^{k-1}
    \end{bmatrix} \right\}  \right\}, \nonumber \\
    &\tilde{\mathcal{J}}_i\left({\mathbf{W}}, \boldsymbol{\eta}, t\right)^{k-1} \triangleq -\mathcal{J}_i\left({\mathbf{W}}, \boldsymbol{\eta}, t\right)^{k-1}- \Real \left\{ \tr \left\{\nabla^H \mathcal{J}_i\left({\mathbf{W}}, \boldsymbol{\eta}, t\right)^{k-1} \begin{bmatrix}
    \lbrace{\mathbf{w}_i-\mathbf{w}_i^{k-1}}\rbrace_{i=1}^{\nuserpercell} \\
    {\eta_{i}-\eta^{k-1}_{i}} \\
    t-t^{k-1}
    \end{bmatrix} \right\}  \right\}. 
    \end{align}
    Following \eqref{eq:Grad}, the expressions for $\nabla\mathcal{H}_i(\mathbf{W}^{k-1},\eta^{k-1}_{i},t^{k-1})$ and $\nabla \mathcal{J}_i(\mathbf{W}^{k-1},\eta^{k-1}_{i},t^{k-1})$ can be obtained.
    Similarly, the first order Taylor series approximation of the objective in $\cal{P}_{6}^{\text{MM}}$ after ignoring the constant terms,\vspace{-0.3cm}
    $$\mathcal{F}\left(t,\boldsymbol{\eta}\right)=t - \lambda_2 \sum_{i=1}^{\nuserpercell}\eta_{i}\nabla \mathbb{P}\left(\eta_{i}^{k-1}\right)+\Omega \left(\sum_{i=1}^{\nuserpercell}\eta_i-\ntxants\right)^2$$
    \item Optimization: The update $\left({\mathbf{W}}^{k}, \boldsymbol{\eta}^{k},t^{k}\right)$ is obtained by solving the following convex problem: \vspace{-0.9cm} 
    \begin{align}
    \cal{P}_{7}^{\text{MM}}:&\text{}\max_{\mathbf{W},\boldsymbol{\eta},t} \hspace{0.10cm} \mathcal{F}\left(t,\boldsymbol{\eta}\right) \nonumber\\  
    \text{s.t. } C_1, &C_2, C_3, C_4 \text{ in } \eqref{eq:Pen_relx_MMSINR_bin} \nonumber \\
    C_5:& \text{ }I_i\left(\mathbf{W},t \right) + \tilde{h}({\mathbf{W}_i^{k-1}},\eta^{k-1}_{i},t^{k-1}) \leq 0, \text{ }\forall i, \nonumber \\
     C_6:& \text{ }l_i\left(\mathbf{W},t \right) + \tilde{h}({\mathbf{W}_i^{k-1}},\eta^{k-1}_{i},t^{k-1}) \leq 0, \text{ }\forall i. \nonumber
    \end{align}
\end{itemize}

 Since, JSP-MMSINR is a CCP based iterative algorithm its complexity depends on the problem $\cal{P}_{7}^{\text{MM}}$. The problem $\cal{P}_{7}^{\text{MM}}$ has $\left(\nuserpercell\ntxants+\nuserpercell+1 \right)$ decision variables, $2\nuserpercell+1$ convex and $2\nuserpercell+1$ linear constraints, hence the computational complexity of $\cal{P}_{7}^{\text{MM}}$ is $\mathcal{O}\left(\left(\nuserpercell \ntxants+\nuserpercell+1 \right)^3 \left(4\nuserpercell+2 \right) \right)$. 
 
\vspace{-0.5cm}
\subsection{Feasible Initial Point: MM-SINR}\label{sec:FIP_MM}
 Unlike WSR problem, obtaining a trivial FIP to the problem $\cal{P}_{12}^{\text{MM}}$ is difficult as initializing $\mathbf{W}$ to all zeros results in zero SINR for all the users and thus $t=0$ where later is the violation of the constraint $C_5$. However, one may find a FIP by the following iterative procedure.
     \begin{itemize}
     \item Step 1: Initialize $\boldsymbol{\eta}=\boldsymbol{\hat{\eta}}$ that satisfies constraints $C_1$ and $C_3$ in $\cal{P}^{\text{MM}}_5$.
      \item Step 2:  For a fixed $\boldsymbol{\eta}$, ignoring the constraints dependent on $t$, $\cal{P}_{6}^{\text{MM}}$ can be reformulated as a convex problem by \cite{SOCP_Amiweisel} or \cite{Bengtsson454064}. Let $\hat{\mathbf{W}}$ be the solution from this step.
   \item Step 3: Exit the loop if $\hat{\mathbf{W}}$ from step 2 is feasible and $t^0 =\dfrac{1}{\min_i \lbrace \eta_i\epsilon_i \rbrace}$ else set $\boldsymbol{{\eta}}=\delta \boldsymbol{\hat{\eta}}$ and continue to step 2.
   \end{itemize}
   \emph{Remarks:}
   \begin{itemize}
     \item The probability of$\hat{\mathbf{W}}$ being feasible increases as ${\hat{\boldsymbol{\eta}}}$ approaches to zero. 
   \end{itemize}

Letting $\cal{P}_{7}^{\text{MM}}\left(k\right)$ be the objective value of the problem $\cal{P}_{7}^{\text{MM}}$ at iteration $k$, the pseudocode of JSP-MMSINR for the joint design problem is given in algorithm~\ref{alg:JSP_MMSINR}. \vspace{-0.1cm}
\begin{algorithm}
 \caption{JSP-MMSINR}
 \label{alg:JSP_MMSINR}
 \begin{algorithmic}[]
 \State \textbf{Input}: $\mathbf{H},\left[\epsilon_1,\hdots,\epsilon_{\nuserpercell}\right],\totpow,\Delta$, $\boldsymbol{\eta}^0$, ${\mathbf{W}}^0$, $\lambda_1=0$, $k=1$
  \State \textbf{Output}: $\mathbf{W},\boldsymbol{\eta}$, $t$
 \While{$|\cal{P}_{7}^{\text{MM}}\left(k\right)-\cal{P}_{7}^{\text{MM}}\left(k-1\right)|\geq \Delta$}
  \State \textbf{Convexification:} Convexify the problem \eqref{eq:MMSINR_convx}
  \State \textbf{Optimization}: Update $\left({\mathbf{W}}^{k}, \boldsymbol{\eta}^{k}, t^k\right)$ by solving $\cal{P}_{7}^{\text{MM}}$ 
  \State \textbf{Update :} $\cal{P}_{7}^{\text{MM}}\left(k\right), \lambda_2, k$;
  \EndWhile
\end{algorithmic}
\end{algorithm}  
\vspace{-0.3cm}
\section{Power Minimization}\label{sec:JSPPMIN}
In this section, we consider the joint design problem with the objective of minimizing the sum power consumed at the BS subject to scheduling of $\ntxants$ users whose minimum SINR requirement is met. As mentioned previously, constraining the Scheduling of utmost $\ntxants$ users leads to the trivial solution of zero users being scheduled whose consumed power is zero.
\vspace{-0.4cm}
\subsection{Joint Design Problem Formulation: PMIN}
Similar to Section~\ref{sec:MMSINR}, the user scheduling is handled through the norm of the precoder as shown in \eqref{eq:SchDef}. With the help of \eqref{eq:SchDef} and notations defined, and letting $\bar{\mathcal{S}}$ to be  the set of scheduled users, a tractable formulation of $\cal{P}_{\text{PMIN}}$ solely as a function of precoding vectors as follows:\vspace{-0.15cm}
\begin{align}\label{eq:PMIN_prob_def}
\cal{P}_{1}^{\text{PMIN}}: &\text{ } \min_{ \mathbf{W},\bar{\mathcal{S}}}  \text{ }\sum_{i \in \bar{\mathcal{S}}} \norm{\mathbf{W}_i}_2^2 \\
\text{s.t. } C_1: &\text{    }\norm{\left[\norm{\mathbf{w}_{1}}_2,\hdots,\norm{\mathbf{w}_{\nuserpercell}}_2\right]}_0 == \ntxants, \text{ }  \nonumber \\    
C_2:& \text{    } 
\gamma_i \geq \Omega_i\epsilon_i, \text{ } i \in \bar{\mathcal{S}}.\nonumber 
\end{align}    
The problem $\cal{P}_{1}^{\text{PMIN}}$ is combinatorial due to the constraints $C_1$ and $C_2$ and also non-convex due to $\lbrace \gamma_i \rbrace_{i=1}^{\nuserpercell}$ in constraint $C_2$. Letting $\Upsilon \in \mathcal{R}^+$ to be a constant, a mathematically tractable formulation that allows us to design a low-complexity algorithm is \vspace{-0.3cm}
\begin{align}\label{eq:PMIN_TracForm}
\cal{P}_{2}^{\text{PMIN}}:&\text{ } \min_{\mathbf{W}, \boldsymbol{\eta}}  \text{ } \norm{\mathbf{W}}_2^2 \\ \text{s.t. }
C_1:& \text{    }\eta_i \in \lbrace 0, 1\rbrace, \text{ } \forall i, \nonumber \\
C_2:& \text{    }\norm{\mathbf{w}_{i}}_2^2 \leq \eta_i \Upsilon, \text{ } \forall i, \nonumber \\
C_3:& \text{    } \sum_{i=1}^{\nuserpercell} \eta_i == \ntxants, \text{ }  \nonumber \\
C_4:& \text{    } \gamma_i \geq {\epsilon}_i \eta_i,\text{ } \forall i. \nonumber \end{align}    
\vspace{-0.2cm}
\emph{Remarks}:
\begin{itemize}
\item For $\eta_i=1$, $\upsilon$ in $C_2$ provides upper bound on the power of user $i$. Moreover, the selection of $\Upsilon$ is trivial as any large $\Upsilon \geq \max_i \text{}\lbrace{\norm{\mathbf{w}_{i}}_2^2}\rbrace$ is valid.
\end{itemize}

\emph{A DC reformulation}: The problem $\cal{P}_{2}^{\text{PMIN}}$ is an MINLP due to combinatorial constraint $C_1$ and non-convex constraint $C_4$. Similar to WSR and MMSINR problems, using the DC formulation of constraint $C_4$ and penalization method for $C_1$, the DC formulation of the problem $\cal{P}_{2}^{\text{PMIN}}$ is, \vspace{-0.3cm}
\begin{align}\label{eq:BinPenPMIN_form}
\cal{P}_{3}^{\text{PMIN}}:&\text{ } \min_{\mathbf{W}, \boldsymbol{\eta}} \text{ } \norm{\mathbf{W}}_2^2 - \lambda_3 \sum_{i=1}^{\nuserpercell} \mathbb{P}\left( \eta_i\right)\\ \text{s.t.  } 
C_1:& \text{    }0 \leq \eta_i \leq  1, \text{ } \forall i, \nonumber \\
C_2, &C_3 \text{ in } \eqref{eq:PMIN_TracForm}, \nonumber \\
C_4:& \text{ }\mathcal{I}_i\left(\mathbf{W}\right)- f_i\left(\mathbf{W}, \eta_i \right) , \text{ } \forall i, \nonumber
\end{align}    
where $\lambda_3 \in \mathcal{R}^{+}$ is the penalty parameter and $f_i\left(\mathbf{W}, \eta_i \right)=\dfrac{\mathcal{I}_i\left(\mathbf{W}\right)+ \lvert \mathbf{h}^H_{i} \mathbf{w}_{i} \rvert^2}{1+{\epsilon}_i\eta_i} $.

The problem $\cal{P}_{3}^{\text{PMIN}}$ is a DC problem which can be solved using CCP. However, finding a FIP becomes difficult as for chosen $\boldsymbol{\eta}$, $\cal{P}_{3}^{\text{PMIN}}$ may become infeasible \cite{SOCP_Amiweisel}. For the ease of finding an FIP, the constraint $C_2$ in $\cal{P}_{4}^{\text{PMIN}}$ is relaxed and penalized as follows: \vspace{-0.3cm}
\begin{align}\label{eq:PMIN_relx_DC_form}
\cal{P}_{4}^{\text{PMIN}}:&\text{} \min_{\mathbf{W}, \boldsymbol{\eta}} \text{}\norm{\mathbf{W}}_2^2 - \Omega\sum_{i=1}^{\nuserpercell} \mathbb{P}\left( \eta_i\right) + \mu \left( \sum_{i=1}^{\nuserpercell} \eta_i-\ntxants\right)^2\\ \text{s.t. } 
C_1,&C_2, C_4 \text{ in } \eqref{eq:BinPenPMIN_form} \nonumber\end{align}    
where $\mu >0$ is penalty parameter. Notice that for the appropriate $\mu$, equality constraint is ensured. Moreover, The problem $\cal{P}_{4}^{\text{PMIN}}$ is a DC problem which solvable using CCP.
\vspace{-0.4cm}
\subsection{Joint Design Algorithm: PMIN}\label{sec:Prop_algo_PMIN}

In this section, following the CCP framework proposed in Section~\ref{sec:Prop_algo_MMSINR}, the CCP based algorithm for PMIN is proposed. The proposed joint scheduling and precoding (JSP) for PMIN (JSP-PMIN) algorithm executes the following two steps iteratively until the convergence:
\begin{itemize}
    \item Convexification: Let ${\mathbf{W}}^{k-1}$, and $\boldsymbol{\eta}^{k-1}$  be the estimates of $\mathbf{W}_{i}$, and $\eta_{i}$ in iteration $k-1$. In iteration $k$, the concave part of $C_3$ in $\cal{P}_{4}^{\text{PMIN}}$ i.e., $-f_i(\mathbf{W},\eta_{i})$ is replaced by its affine approximation around the estimate of  $\left({\mathbf{W}}^{k-1}, \boldsymbol{\eta}^{k-1} \right)$ which is given by, \vspace{-0.3cm}
    \begin{align}
        &\tilde{f}(\mathbf{W},\eta_{i};{\mathbf{W}_i^{k-1}},\eta^{k-1}_{i}) \triangleq -f({\mathbf{W}^{k-1}},\eta^{k-1}_{i})- \Real \left\{ \tr \left\{\nabla^H f(\mathbf{W}^{k-1},\eta^{k-1}_{i})\begin{bmatrix}
    {\mathbf{w}_1-\mathbf{w}_1^{k-1}} \\
    \vdots \\
    {\mathbf{w}_{\nuserpercell}-\mathbf{w}_{\nuserpercell}^{k-1}} \\
    {\eta_{i}-\eta^{k-1}_{i}}
    \end{bmatrix} \right\}  \right\}.\label{eq:PMIN_convx}
    \end{align}
    \item Optimization: Update $\left({\mathbf{W}}^{k},\pow^{k}, \boldsymbol{\eta}^{k}\right)$ is obtained by solving the following convex problem: \vspace{-0.3cm}
    \begin{align}\label{eq:PMIN_Convexified_CCP}
\cal{P}_{5}^{\text{PMIN}}:&\min_{\mathbf{W}, \boldsymbol{\eta}} \text{}\norm{\mathbf{W}}_2^2+\mu \left(\sum_{i=1}^{\nuserpercell}\eta_i-\ntxants\right)^2 - \lambda_3 \sum_{i=1}^{\nuserpercell}\eta_{i}\nabla \mathbb{P}\left(\eta_{i}^{k-1}\right) \\ 
\text{s.t. } 
C_1:& \text{    } 0 \leq \eta_i \leq 1, \text{ } \forall i, \nonumber \\
C_2:& \text{  } \norm{\mathbf{w}_i}_2^2 \leq  \eta_i \Upsilon, \forall i, \nonumber \\
C_3:& \text{    }\mathcal{I}_i\left(\mathbf{W}\right)+\tilde{f}({\mathbf{W}_i^{k-1}},\eta^{k-1}_{i})  \leq 0, \text{ }\forall i. \nonumber 
\end{align}
\end{itemize}

The convex problem $\cal{P}_{5}^{\text{PMIN}}$ has $\left(\nuserpercell\ntxants+\nuserpercell \right)$ decision variables and $2\nuserpercell$ convex and $2\nuserpercell$ linear constraints, hence the computational complexity of $\cal{P}_{5}^{\text{MM}}$ is $\mathcal{O}\left(\left(\nuserpercell \ntxants+\nuserpercell \right)^3 \left(4\nuserpercell\right) \right).$

\vspace{-0.5cm}
\subsection{Feasible Initial Point: PMIN}
 An initial feasible point for the problem $\cal{P}_{5}^{\text{PMIN}}$ is obtained by the following iterative procedure.
     \begin{itemize}
     \item Step 1: Initialize $\boldsymbol{\eta}=\boldsymbol{\hat{\eta}}$ that satisfies  $C_1$ and $C_3$ in $\cal{P}_{4}^{\text{PMIN}}$.
     \item Step 2:  The precoding problem of $\cal{P}_{4}^{\text{PMIN}}$ for fixed $\boldsymbol{\eta}$ can be reformulated as a convex problem by \cite{SOCP_Amiweisel} or \cite{Bengtsson454064}. Let $\hat{\mathbf{W}}$ be the solution from this step.
   \item Step 3: Exit the loop if $\hat{\mathbf{W}}$ is feasible (see \cite{SOCP_Amiweisel}) else set $\boldsymbol{\eta} = \delta \boldsymbol{\hat{\eta}}$ and continue to step 2.
   \end{itemize}

Letting $\cal{P}_{5}^{\text{PMIN}}\left( k \right)$ be the objective value of the problem $\cal{P}_{5}^{\text{PMIN}}$ at iteration $k$, The pseudo code of the algorithm is illustrated in the table~\ref{alg:JSP_PMIN}. 
\vspace{-0.2cm}
 \begin{algorithm}
  \caption{JSP-PMIN}
  \label{alg:JSP_PMIN}
 \begin{algorithmic}[]
 \State  \textbf{Input}: $\mathbf{H},\left[\bar{\epsilon}_1,\hdots,\bar{\epsilon}_{\nuserpercell}\right],\Delta$, $\boldsymbol{\eta}^0$, ${\mathbf{W}}^0$, $\lambda_1=0$, $k=1$
 \State  \textbf{Output}:$\mathbf{W},\boldsymbol{\eta}$
 \While{$|\cal{P}_{6}^{\text{PMIN}}\left(k\right)-\cal{P}_{6}^{\text{PMIN}}\left(k-1\right)|\geq \Delta$}
  \State \textbf{Convexification:} Convexify the problem \eqref{eq:MMSINR_convx}
  \State \textbf{Optimization}: Update $\left({\mathbf{W}}^{k}, \boldsymbol{\eta}^{k}\right)$ by solving $\cal{P}_{6}^{\text{PMIN}}$ 
  \State \textbf{Update :} $\cal{P}_{6}^{\text{PMIN}}\left(k\right), \Omega, k$
  \EndWhile
 \end{algorithmic}
\end{algorithm}

\section{Simulation results}\label{sec:simulation}

\subsection{Simulation Setup}\label{sec:SimSet}
    In this section, we evaluate the performance of the proposed algorithms for the MMSINR, WSR and PMIN problems. The system parameters and benchmark scheduling method discussed in this paragraph are common for all the figures. Entries of the channel matrix, i.e., $\{h_{ij}\}$s are drawn from the complex normal distribution with zero mean and unit variance and noise variances are considered to be unity i.e., $\sigma^2=1,$ $\forall i$. Simulation results in all the figures are averaged over 500 different CRs.  
    The penalty parameter $\lambda_1$ is initialized to 0.5 and incremented as $\lambda_1=1.1\lambda_1$ until $\lambda_1 \leq 10$. By the nature of MMSINR (PMIN) design, dropping the user with lowest SINR (higher power) leads to the better objective. This phenomenon continues until it drops $\nuserpercell-\ntxants$ users and can not drop any further due to the scheduling constraint. Since, this naturally enforces the binary nature of $\boldsymbol{\eta}$, $\lambda_2=0$ ($\lambda_3=0$) in MMSINR (PMIN) still yields the binary $\boldsymbol{\eta}$ which is shown Section~\ref{sec:PerfMMSINR} and \ref{sec:PerfPMIN}. Hence, $\lambda_2$ and $\lambda_3$ are fixed zero in all iterations. The penalty parameters  $\Omega$ and $\mu$ are initialized to 0.01 and incremented as $\Omega=1.2\Omega$ and $\mu=1.2\mu$ in each iteration until $\Omega \leq 20$ and $\mu \leq 20$.
    
     To evaluate the performance of the proposed JSP algorithms - due to the lack of a comparable joint solution - the following benchmarks (iterative decoupled solutions that execute the following steps in sequence) are devised:
        \begin{itemize}
            \item  In step 1, users are scheduled according to proposed weighted semi-orthogonal user scheduling (WSUS). The considered WSUS is an extension of the SUS algorithm proposed in \cite{ChanOrthoZFBF_goldsmith}. In SUS, the users are selected sequentially based on the channel orthogonality of the scheduled users with yet to be scheduled users channels. In WSUS, orthogonality indices calculated according to SUS are multiplied with its associated weights and the user with the highest weighted orthogonality index is scheduled. This process is repeated until $\ntxants$ users are scheduled. 
            \item In step 2, the precoding problem for the  scheduled users is solved by the following methods:
            \begin{itemize}
                \item 
            It is easy to see that, keeping only the terms corresponding to scheduled users and substituting corresponding $\eta_i$s to 1 and ignoring the constraint solely dependent on $\eta_i$s in \eqref{eq:Penalized_WSR_bin} and  \eqref{eq:Pen_relx_MMSINR_bin}  
            gives the DC formulation of the precoding problem for the scheduled users for WSR and MMSINR and respectively. These precoding problems can be solved using CCP with a FIP  obtained from \ref{sec:FIP_WSR} and \ref{sec:FIP_MM} by substituting corresponding $\eta_i$s with 1. SUS and WSUS combined with this proposed WSR is simply referred to as SUS-WSR and WSUS-WSR respectively and for MMSINR as SUS-MMSINR and WSUS-MMSINR respectively. The SDP based power minimization proposed in \cite{Bengtsson454064} is used for PMIN precoding problem and is referred to simply as SUS-PMIN and WSUS-PMIN for the users scheduled based on SUS and WSUS respectively.
            \item An SDR version of DC formulation proposed in \cite{CC_2017} also used for solving the precoding for the scheduled users in WSR case as a reference hence is referred to as RWSR. WSUS combined with RWSR is referred to as WSUS-RWSR.
            \end{itemize}
            \item In step 3: If the precoding problem in step 2 is infeasible exit the loop else drop the user with least orthogonality and repeat step 2 for an updated set of scheduled users. However, the precoding problems for MMSINR and PMIN are assumed to be feasible.
            \end{itemize}
    
    \vspace{-0.6cm}
    \subsection{WSR Performance Evaluation}\label{subsec:PerfWSR}    
\begin{figure}
	\centering
	\includegraphics[height=6.5cm,width=10cm]{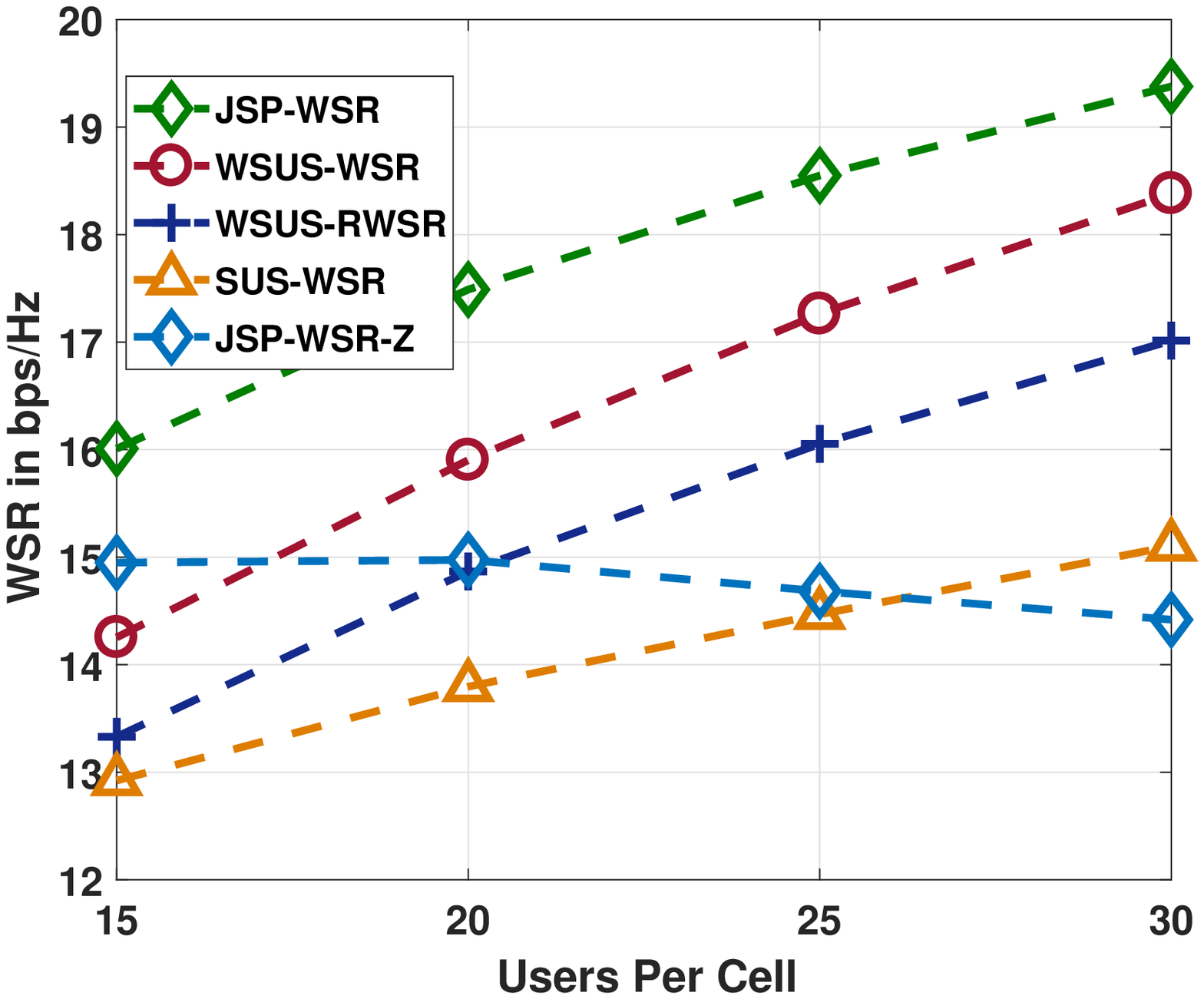}
	\caption{Performance comparison of WSR for $\ntxants=$10, $\lbrace \epsilon_i =4\text{dB}\rbrace_{i=1}^{\nuserpercell},$ $\totpow$=10 dB, and $\nuserpercell$ is varied from 15 to 30 insteps of 5.}\label{fig:WSR_vs_N}	
\end{figure}
     In figure \ref{fig:WSR_vs_N}, we compare the performance of JSP-WSR as a function of $N$ varying from 15 to 30 in steps of 5 for $M=10$, $\totpow=10$dB and $\epsilon_i=4\text{dB}$, $\forall i$. Weights $\lbrace \alpha_i\rbrace_{i=1}^{\nuserpercell}$ are randomly drawn from the set $\{\frac{k}{N}\}, k=1, \ldots, N$. In figure \ref{fig:WSR_vs_N}, SUS-WSR, WSUS-WSR and WSUS-RWSR are the decoupled benchmark algorithms. The JSP-WSR initialized with a trivial solution ($\mathbf{W}^0=\mathbf{0}$, $\boldsymbol{\eta}^0=\mathbf{0}$) is referred to as JSP-WSR-Z and JSP-WSR initialized with an FIP obtained from Section~\ref{sec:FIP_WSR} continues to be referred to as JSP-WSR. 
     
     From figure~\ref{fig:WSR_vs_N}, it is clear that the joint solution JSP-WSR outperforms all the other decoupled benchmarks.  Although JSP-WSR, SUS-WSR, and WSUS-WSR have the same underlying precoding algorithm, JSP-WSR achieves better performance as it jointly updates scheduling and precoding. Considering weights into scheduling in WSUS-WSR improves over SUS-WSR, as shown in figure~\ref{fig:WSR_vs_N}, it still underperforms compared to JSP-WSR. However, the gains diminish as $\nuserpercell$ increases as the probability of finding near orthogonal channels increases which means scheduling the users with negligible interference. Hence, WSUS-WSR performs close to JSP-WSR for $\nuserpercell$  relatively larger than $\ntxants$. Notice that despite the difference in the rate of growth, all methods improve SR as $\nuserpercell$ increases due to multiuser diversity.
     
    Notice that JSP-WSR and JSP-WSR-Z are identical except the FIPs. JSP-WSR and JSP-WSR-Z are CCP based algorithms hence the performance differentiation depends on FIP. Figure~\ref{fig:WSR_vs_N} shows that while a poor FIP like $\mathbf{W}^0=\mathbf{0}$, $\boldsymbol{\eta}^0=\mathbf{0}$ results in worse performance than decoupled solutions, the FIPs from Section~\ref{sec:FIP_WSR} achieves better performance. This shows that FIPs obtained from \ref{sec:FIP_WSR} are generally good. Particularly, $\mathbf{W}^0=\mathbf{0}$, $\boldsymbol{\eta}^0=\mathbf{0}$ is a bad choice since it is the solution that achieves lowest WSR i.e., zero and hence the solutions of JSP-WSR-Z are generally the stationary points around the lowest objective.  
    
    Despite having the same WSUS scheduling algorithm and the same FIP for precoding, WSUS-WSR outperforms WSUS-RWSR due to the difference in precoding algorithms as shown in figure~\ref{fig:WSR_vs_N}. Although WSRP can be formulated as a DC problem using proposed reformulations and also by the approach in \cite{CC_2017}, due to the efficiency of proposed reformulations, WSUS-WSR achieves the better objective which is confirmed by figure~\ref{fig:WSR_vs_N}.
    
\begin{figure}
    	\centering
    	\subfloat[]{\label{fig:SR_vs_N} \includegraphics[height=5cm,width=6cm]{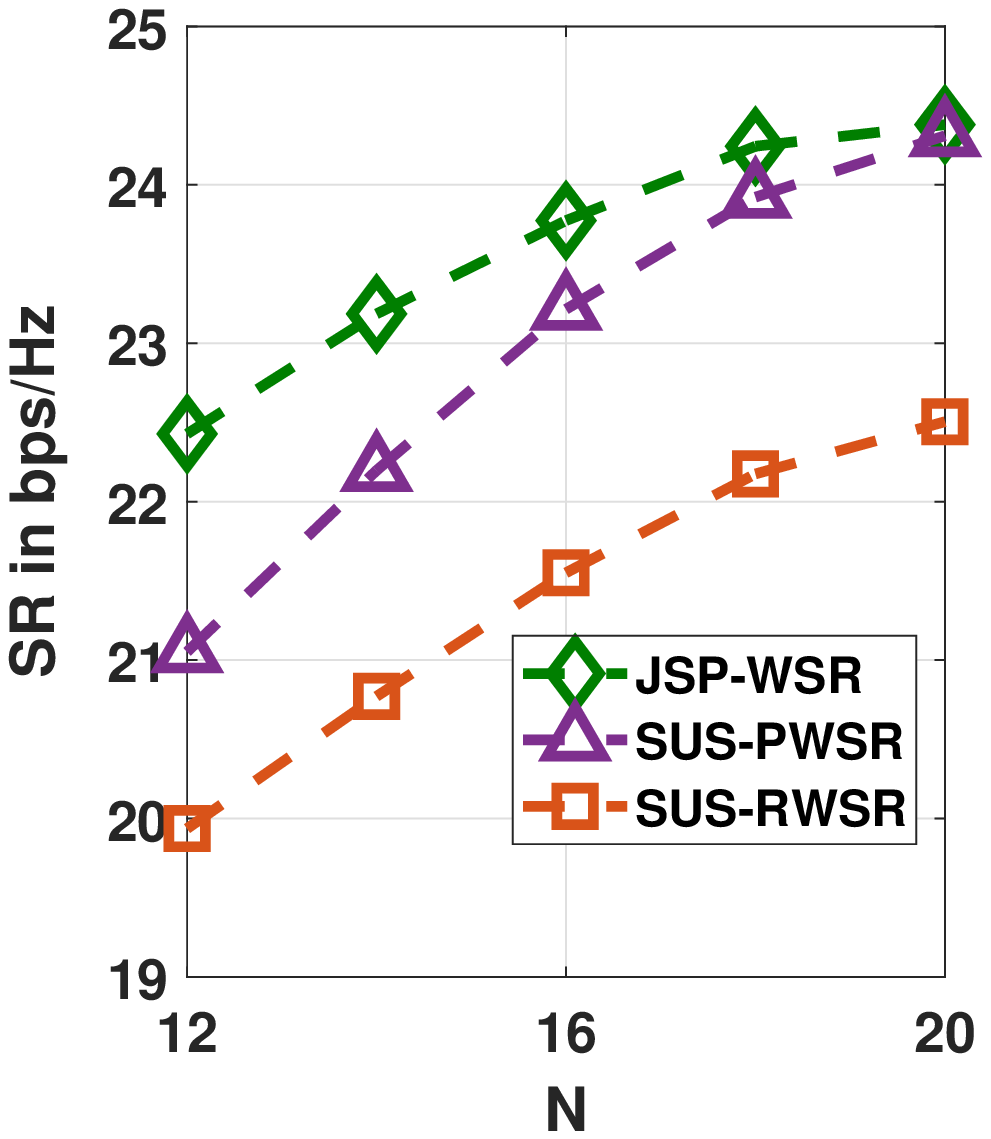} }
    	\subfloat[]{\includegraphics[height=5cm,width=6cm]{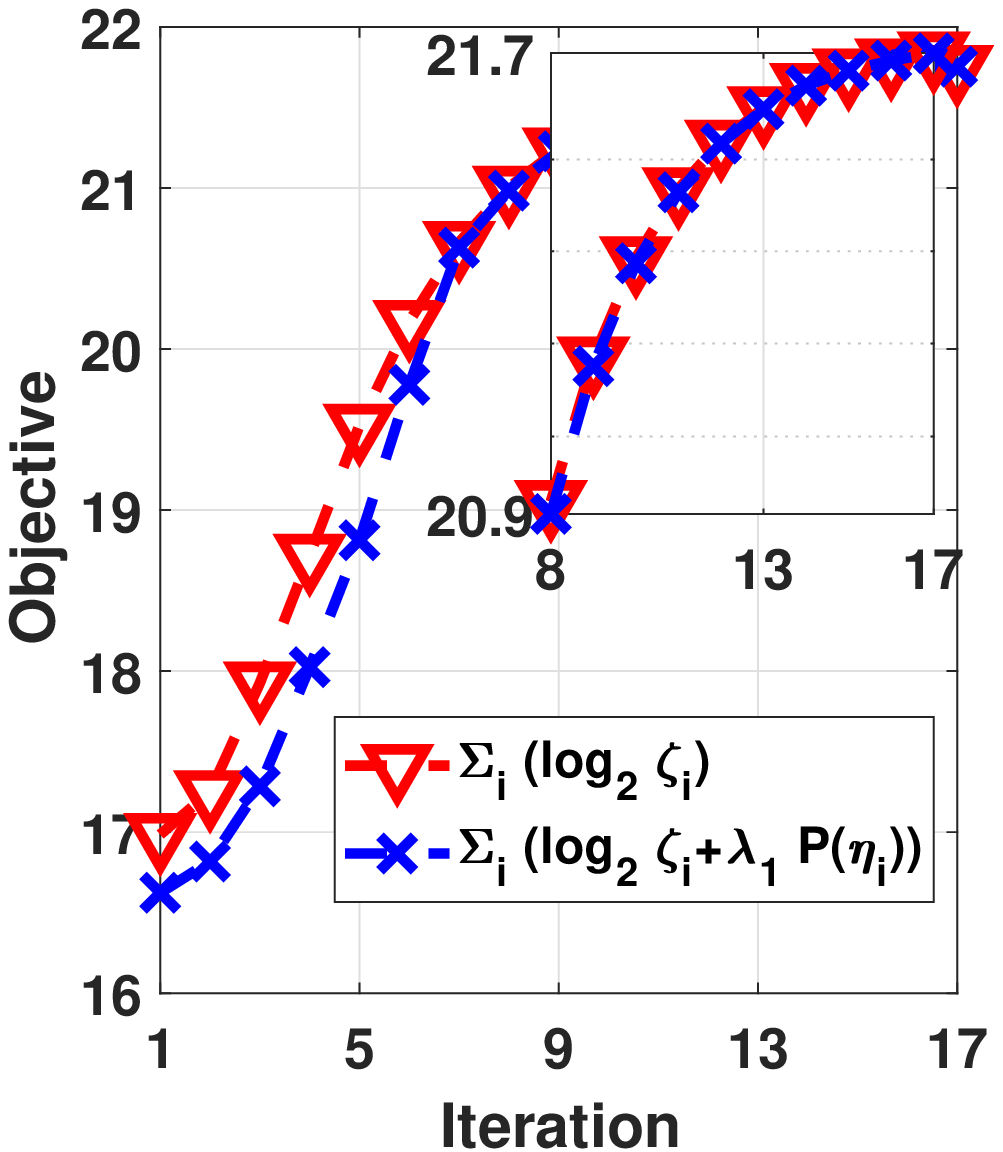}\label{fig:ObjVal_WSR}}
    	\caption{(a) Performance comparison of SR for uniform weighted case with  $\ntxants=$10, $\lbrace \epsilon_i =4\text{dB}\rbrace_{i=1}^{\nuserpercell}$, $\totpow=10\text{dB}$ and $\nuserpercell$ varying from 12 to 20 in steps of 2. (b) Illustration of convergence of the JSP-WSR (with penalty) and convergence of $\boldsymbol{\eta}$ to binary for $\ntxants=$10, $\lbrace \epsilon_i =4\text{dB}\rbrace_{i=1}^{\nuserpercell}$, $\totpow=10\text{dB}$ and $\nuserpercell=20$. }\label{fig:WSR_perf}
    \end{figure}

    
    The performance of the JSP-WSR is illustrated for uniform weighted case i.e. $\lbrace \alpha_i=1\rbrace_{i=1}^{\nuserpercell}$ in figure~\ref{fig:SR_vs_N} as a function of $N$. The performance gain by jointly updating scheduling and precoding in JSP-WSR over the decoupled SUS-WSR and SUS-RWSR is clear from figure~\ref{fig:SR_vs_N}. However, as $\nuserpercell$ increases ($\nuserpercell \approx 20$) SUS schedules the users with strong channel gains and least interference hence SUS-WSR performs close to JSP-WSR. Despite the efficiency of SUS in the region around $\nuserpercell=20$, SUS-RWSR performs poor due to the inefficiency of the RWSR precoding scheme.
    
    In figure~\ref{fig:ObjVal_WSR}, the convergence behavior of the JSP-WSR and the convergence of $\boldsymbol{\eta}$ to binary values is illustrated as a function of iterations. The SR obtained in each iteration is shown by the red curve while the penalized SR is shown by the blue curve. As the FIP of JSP-WSR contains a non-binary $\boldsymbol{\eta}$, the solutions obtained in the initial iterations include the non-binary $\boldsymbol{\eta}$; hence, the difference between SR (red curve) and SR plus penalty (blue curve). However, as the penalty factor ($\lambda_1$) increases over the iterations, JSP-WSR favors the solutions with $\eta_i$s close to 0 or 1, hence over the iterations penalty approaches zero i.e., $ \mathbb{P}\left(\eta_{i}\right)\approx0, \text{}\forall i$. This behavior is clear from iteration 8 onwards. Moreover, the convergence behavior of the JSP-WSR to a stationary point of $\cal{P}_5^{\text{WSR}}$ is shown by the convergence of the blue curve which depicts its objective value.

    \vspace{-0.5cm}
        \subsection{MMSINR Performance Evaluation }\label{sec:PerfMMSINR}
        

     \begin{figure}
     \centering
        \subfloat[]{\includegraphics[height=5cm,width=6cm]{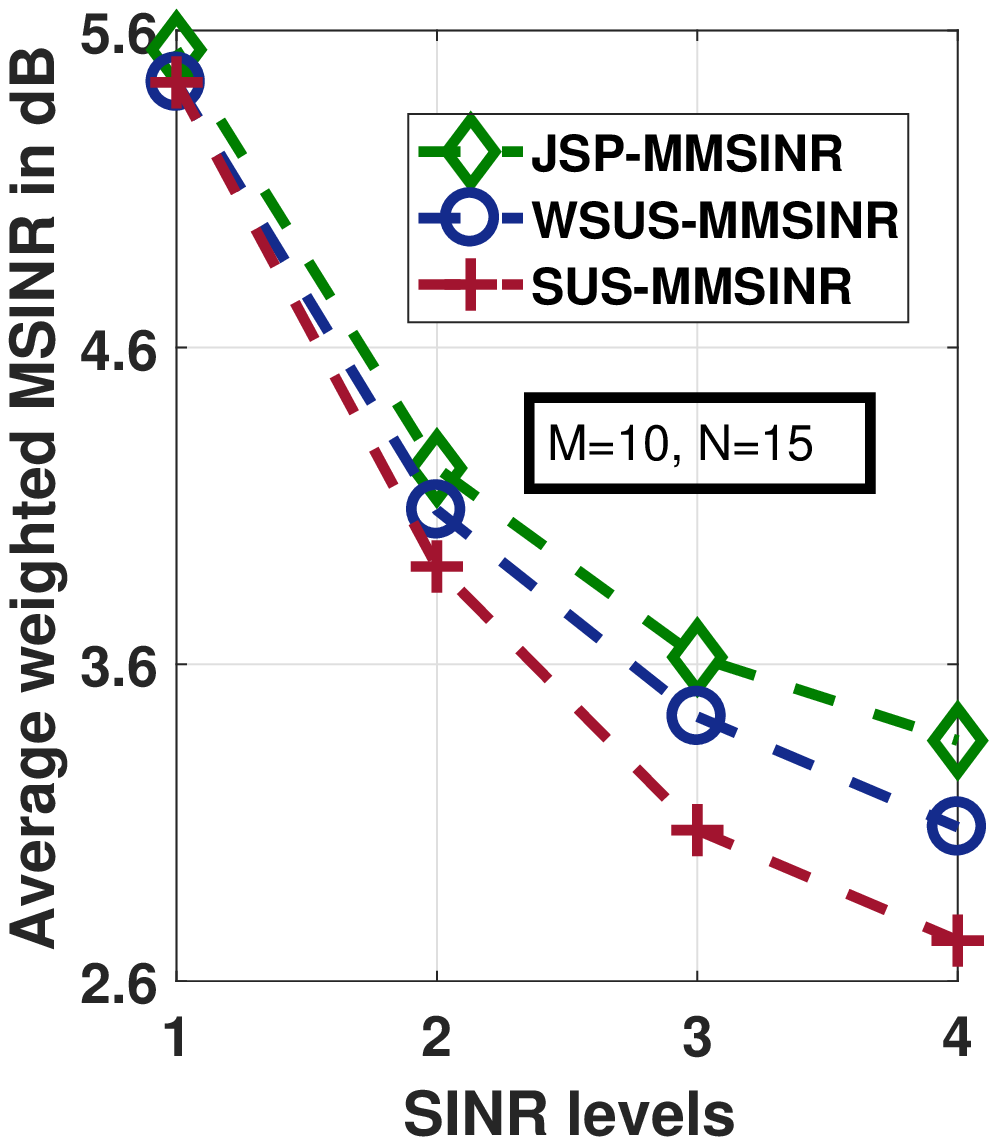}\label{fig:MinSINRvsSINRLevs_N15}}
        \subfloat[]{\includegraphics[height=5cm,width=6cm]{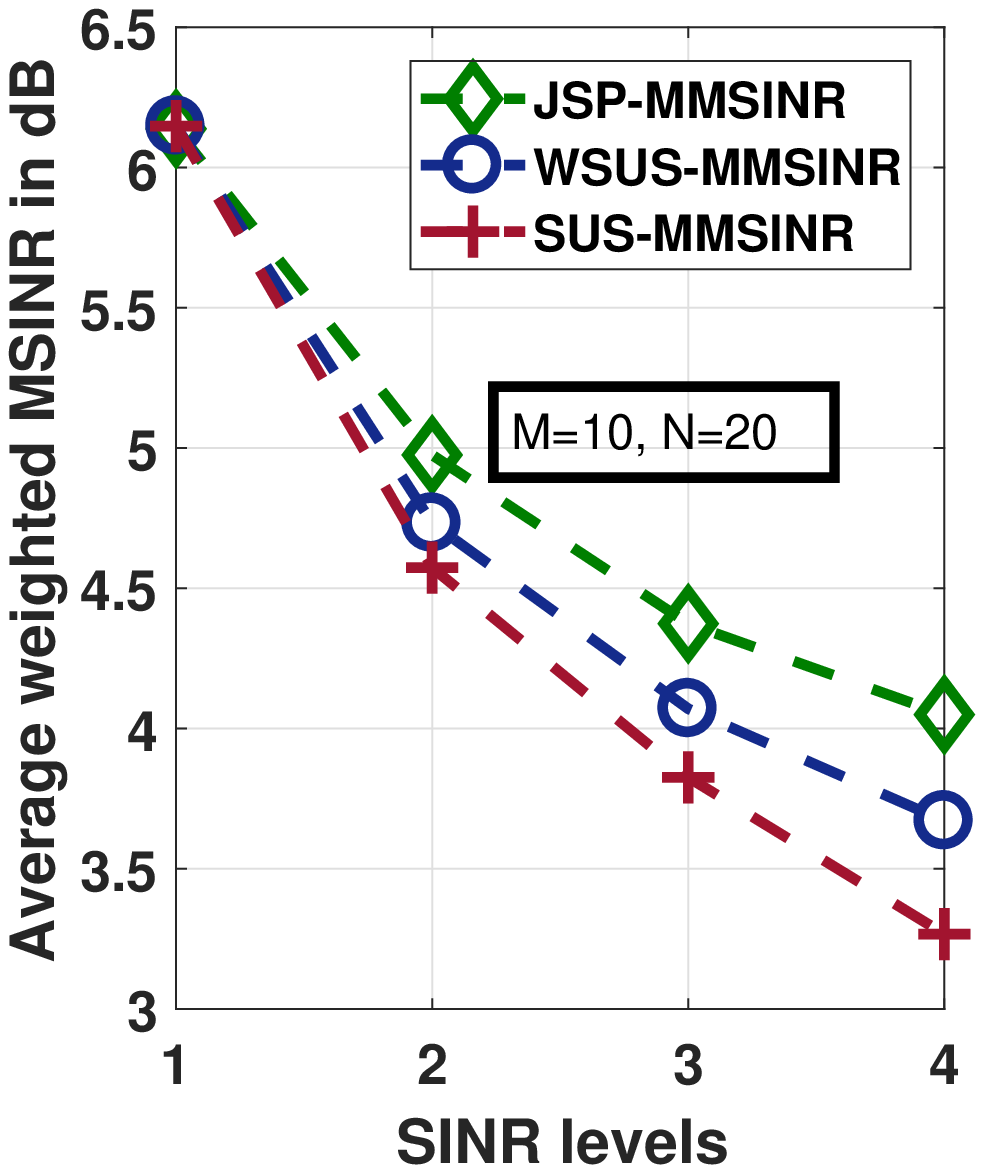}\label{fig:MinSINRvsSINRLevs_N20}}
        \caption{Performance comparision of of MMSINR versus SINR levels for $\totpow=10$dB, $\lbrace \epsilon_i=0\text{dB}\rbrace_{i=1}^{\nuserpercell}$ with $\ntxants=15$ in (a) and $\ntxants$=20 in (b).}\label{fig:MMSINRvsSINRLevs}
    \end{figure}
    \begin{figure}
        \centering
        \subfloat[]{\includegraphics[height=5cm,width=6cm]{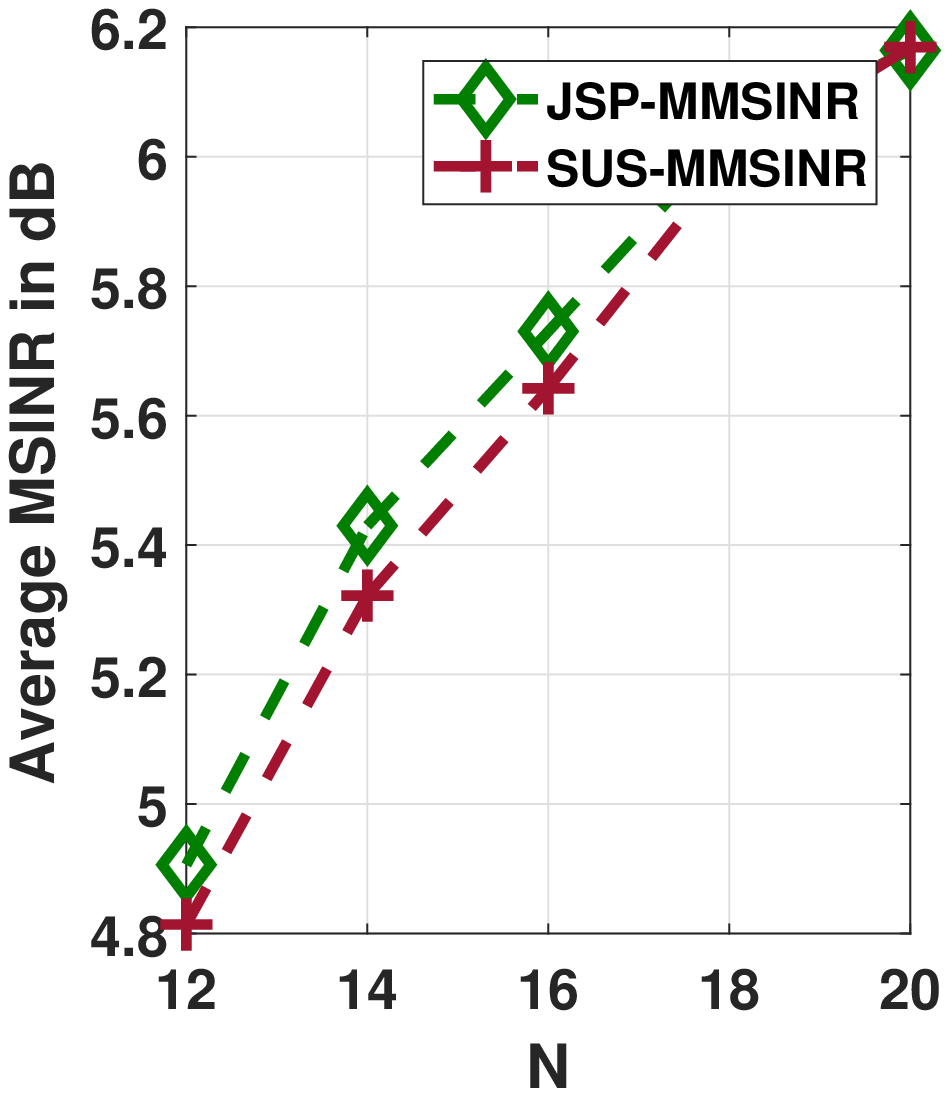}\label{fig:MMSINRUnweighted}}
        \subfloat[]{\includegraphics[height=5cm,width=6cm]{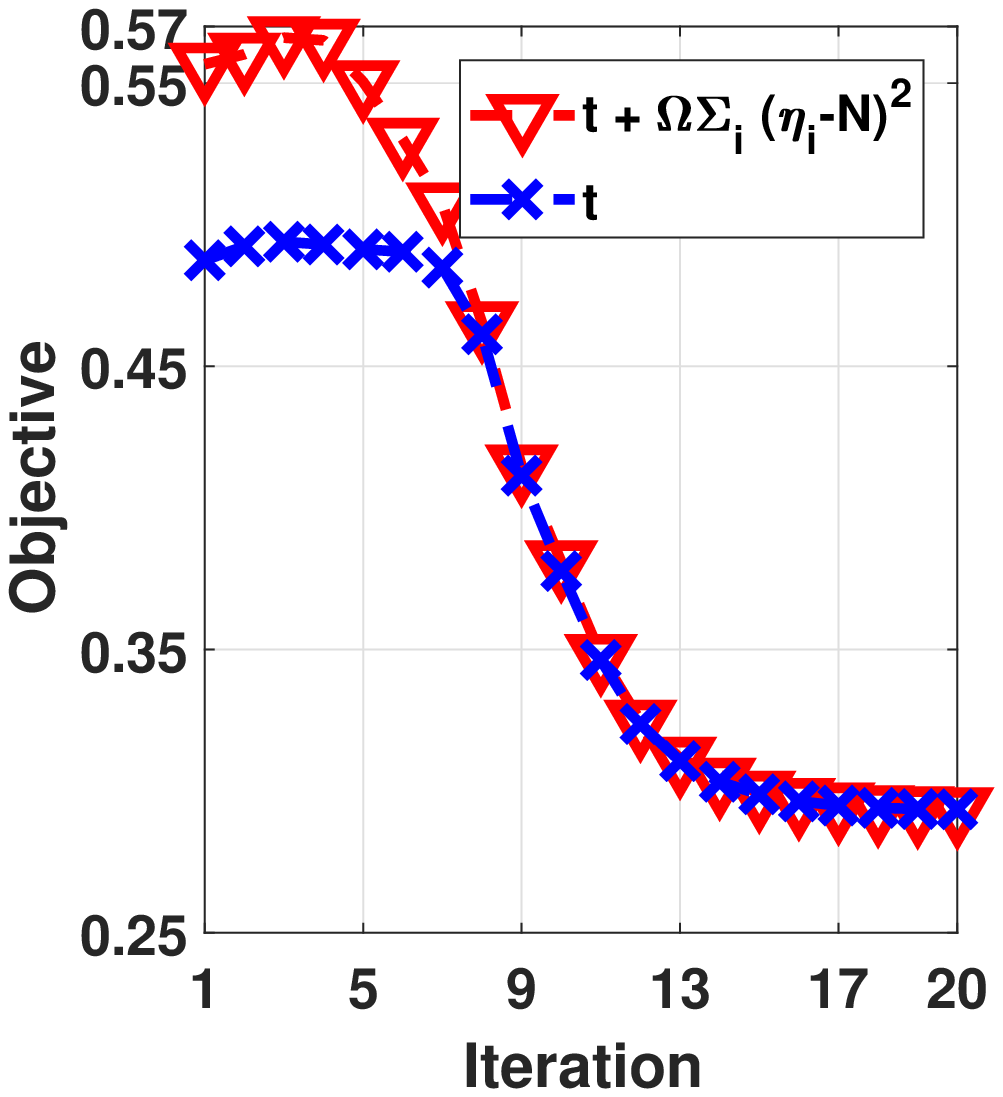} \label{fig:ObjValMMSINR}}
        \caption{(a) Performance comparison of MMSINR for uniform weighted case with  $\ntxants=$10,$\lbrace \beta_i=1, \epsilon_i=0\text{dB}\rbrace_{i=1}^{\nuserpercell}$, $\totpow$=10dB and $\nuserpercell$ varying from 12 to 20 in steps of 2. (b) Illustration of convergence of the JSP-MMSINR (with penalty) and convergence of $\boldsymbol{\eta}$ to binary for  $\ntxants=$10,$\lbrace \beta_i=1, \epsilon_i=0\text{dB}\rbrace_{i=1}^{\nuserpercell}$, $\totpow$=10dB and $\nuserpercell=20$.}\label{fig:PerfMMSINR}
    \end{figure}
    
    In figure~\ref{fig:MMSINRvsSINRLevs}, the performance of JSP-MMSINR is compared with SUS-MMSINR and WSUS-MMSINR for $\ntxants=10$, $\lbrace\epsilon_i=1\text{ (0 dB) }\rbrace_{i=1}^{\nuserpercell}$, $\totpow=10\text{dB}$ and $\nuserpercell=15$ in figure~\ref{fig:MinSINRvsSINRLevs_N15} and $\nuserpercell=20$ in figure~\ref{fig:MinSINRvsSINRLevs_N20}. In figure~\ref{fig:MMSINRvsSINRLevs}, the weighted minimum SINR (MSINR) of the scheduled users is averaged over 500 different CRs is referred to as average weighted MSINR and is illustrated as a function of SINR levels. For SINR level 1, 2, 3 and 4, the weight $\beta_i$ associated with user $i$ is randomly drawn from the sets $\{1\}$, $\{0.5,1\}$, $\{0.333, 0.6666, 0.9999\}$ and $\{0.25, 0.5, 0.75, 1\}$ respectively. For example, for SINR levels 2, ${\beta_i}$ is randomly selected from $\{{0.5}, 1\}$. Hence the MMSINR requirement of each user is 1/0.5 or 1 since $\epsilon_i=1$.  Notice that a higher value of $\beta_i$ increases the likeliness of user $i$ being scheduled. It is clear from figure~\ref{fig:MMSINRvsSINRLevs}, that the joint solution JSP-MMSINR improves the performance over the decoupled design SUS-MMSINR and WSUS-MMSINR. Despite the identical underlying precoding scheme in JSP-MMSINR, SUS-MMSINR, and WSUS-MMSINR, the systematic joint update of scheduling and precoding considering the weights is helping JSP-MMSINR to achieve better performance. Although WSUS-MMSINR achieves better performance over SUS-MMSINR by considering the weights into scheduling, it still performs worse than JSP-MMSINR showing the inefficiency of decoupled design. 
    
    The performance of JSP-MMSINR is illustrated for uniform weighted case i.e., $\lbrace \beta_i=1\rbrace_{i=1}^{\nuserpercell}$ in figure~\ref{fig:PerfMMSINR} for $\ntxants=10$ and $\totpow=10\text{dB}$. In figure~\ref{fig:MMSINRUnweighted}, the average MSINR is illustrated as a function of $\nuserpercell$ varying from 12 to 18 in steps of 2. The superior performance of JSP-MMSINR over SUS-MMSINR is clear from \ref{fig:MMSINRUnweighted}. However, the gains diminish as $\nuserpercell$ increases as the SUS based solution becomes efficient as mentioned previously. In figure~\ref{fig:ObjValMMSINR}, the convergence behavior of the algorithm and progression of achieving exact scheduling constraint i.e., $\sum_{i=1}^{\nuserpercell}==\ntxants$ as function of iteration is illustrated. While the blue curve depicts the inverse of MSINR achieved over the iteration, the red curve depicts the penalized objective where the penalty is for ensuring the constraint of scheduling exactly $\ntxants$ users. As FIPs violate the exact scheduling constraint, the penalized objective (red curve) is far from the objective (blue curve). However, increasing the penalty parameter $\Omega$ over the iterations until $\Omega\leq 20$ ensures the scheduling constraint. This behavior is observed from iteration 8 in figure~\ref{fig:ObjValMMSINR} as the difference between penalized objective and objective is approximately zero. Moreover, the binary nature of $\boldsymbol{\eta}$ is also achieved over the iterations due to nature of MMSINR for fixed $\lambda_2=0$ in figure~\ref{fig:MMSINRvsSINRLevs} and \ref{fig:PerfMMSINR}.
    
    \vspace{-0.4cm}
     \subsection{PMIN Performance Evaluation }\label{sec:PerfPMIN}
\begin{figure}
\centering
        \subfloat[]{\includegraphics[height=5.0cm,width=6cm]{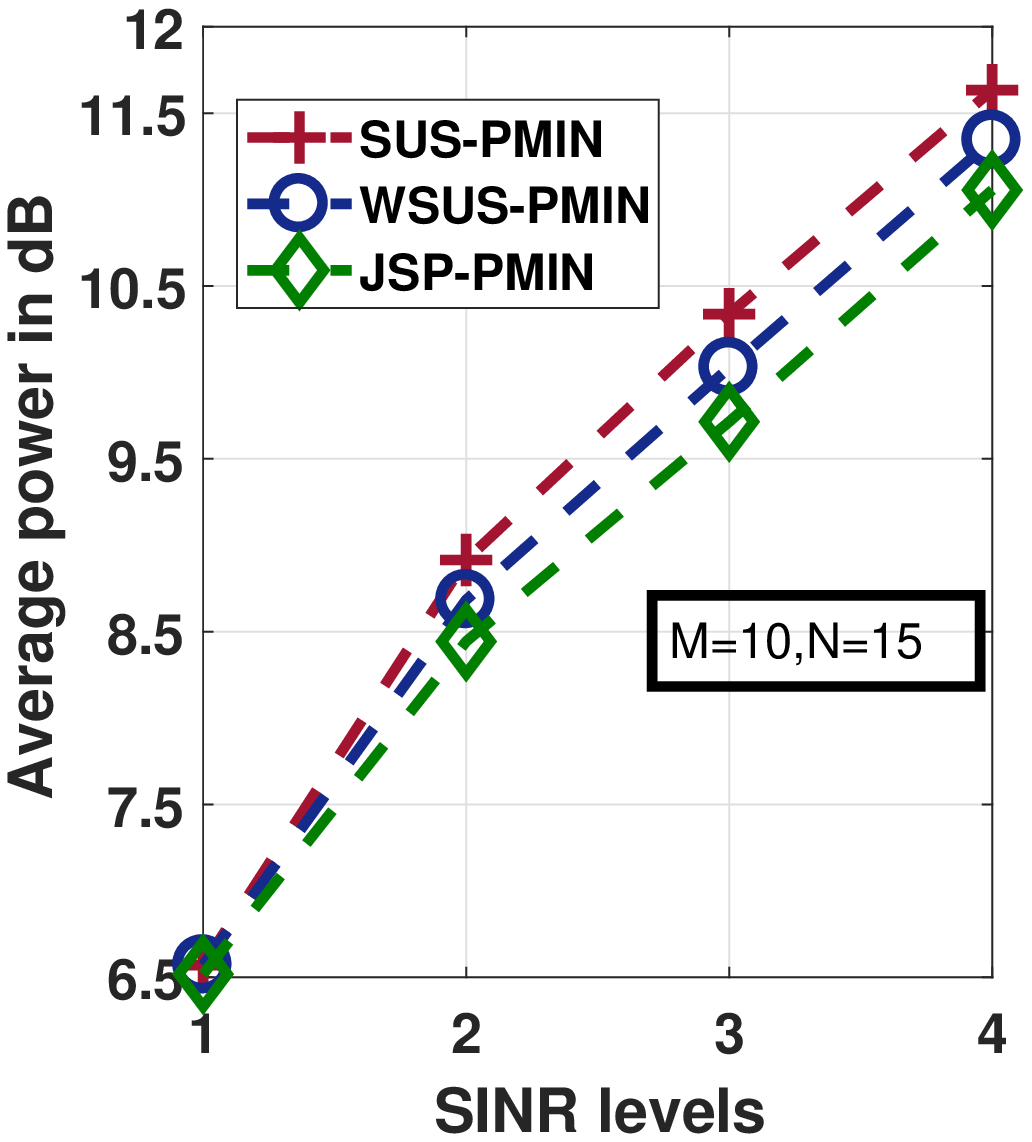}\label{fig:PMIN_N15}}
        \subfloat{\includegraphics[height=5.15cm,width=6cm]{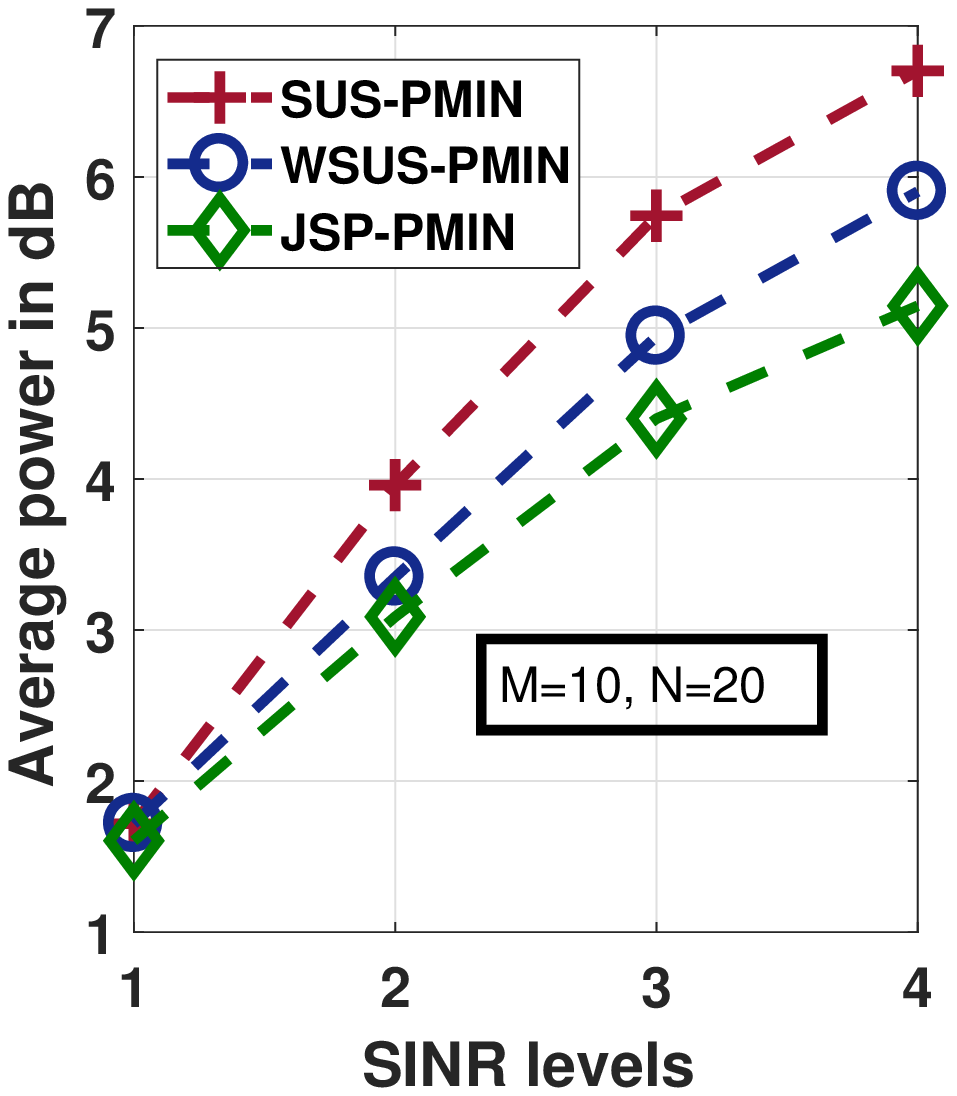}\label{fig:PMIN_N20}}
     \caption{Performance comparision of PMIN versus SINR levels for $\ntxants=10$ and $\totpow=10$dB. }\label{fig:PMIN_comparision} 
\end{figure}

    The total power consumed by the scheduled users (for each CR) is averaged over 500 CRs which is referred to as average total power per CR. In figure~\ref{fig:PMIN_comparision}, the average total power per CR is depicted as a function of SINR levels for $\ntxants=10$, $\nuserpercell=15$ in figure~\ref{fig:PMIN_N15} and $\nuserpercell=20$ in figure~\ref{fig:PMIN_N20}. The SINR level 1, 2, 3 and 4 (different than MMSINR design) on the x-axis indicate that ${\epsilon}_i$ is randomly chosen from the sets $\lbrace 1 \rbrace$,  $ \lbrace 1, 2 \rbrace$,  $\lbrace 1, 2, 3 \rbrace$ and  $\lbrace 1, 2, 3, 4\rbrace$ for user $i$ respectively. For example, for the SINR level 2, ${\epsilon}_i$ for user $i$ is randomly chosen from the set $ \lbrace 1, 2 \rbrace$. It is clear from figure~\ref{fig:PMIN_N15} and  \ref{fig:PMIN_N20}, that the joint solution JSP-PMIN outperforms SUS-PMIN and WSUS-PMIN. Although the precoding problem for the scheduled users by SUS and WSUS is solved globally using \cite{Bengtsson454064}, the inefficient scheduling leads to the poorer performance over JSP-PMIN. 
    
\begin{figure}
    \centering
\subfloat[]{\includegraphics[height=5cm,width=6cm]{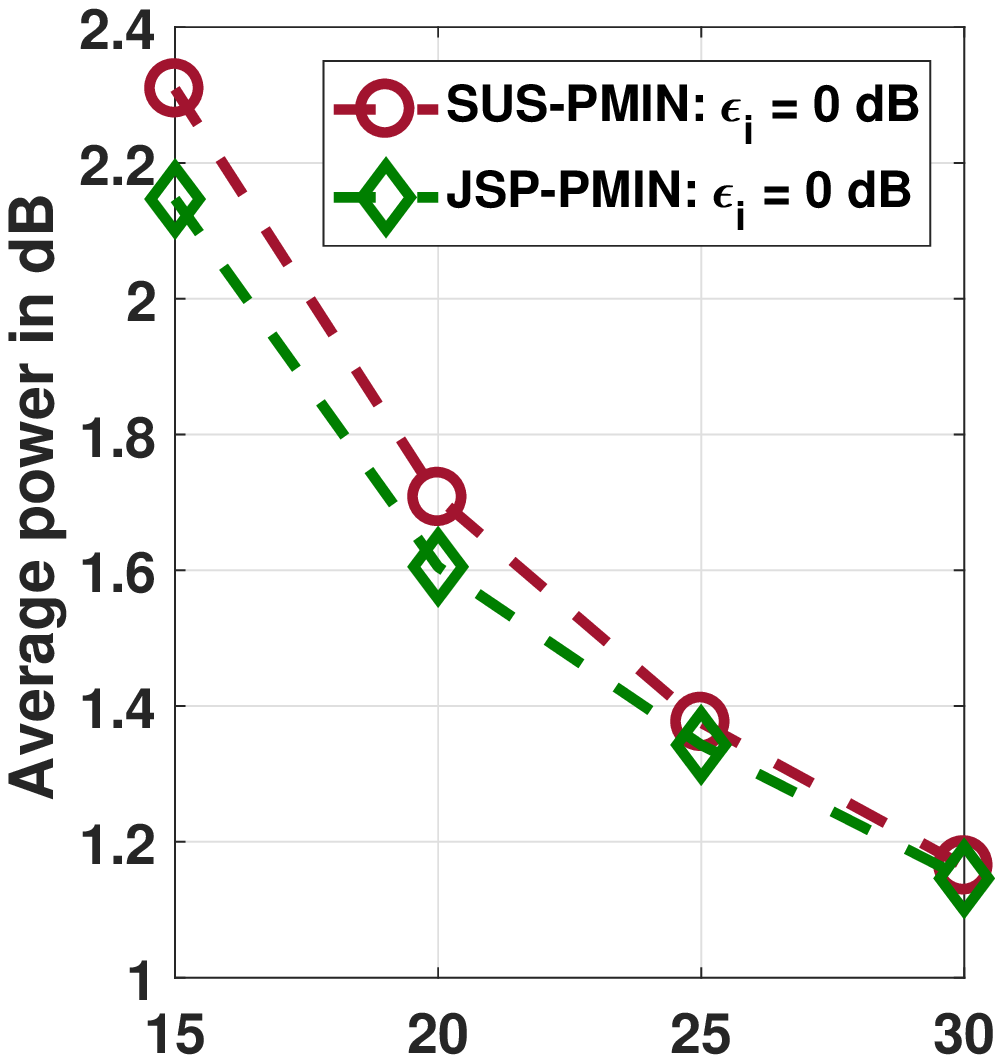}\label{fig:PMINUnwegihted}}\subfloat[]{\includegraphics[height=5cm,width=6cm]{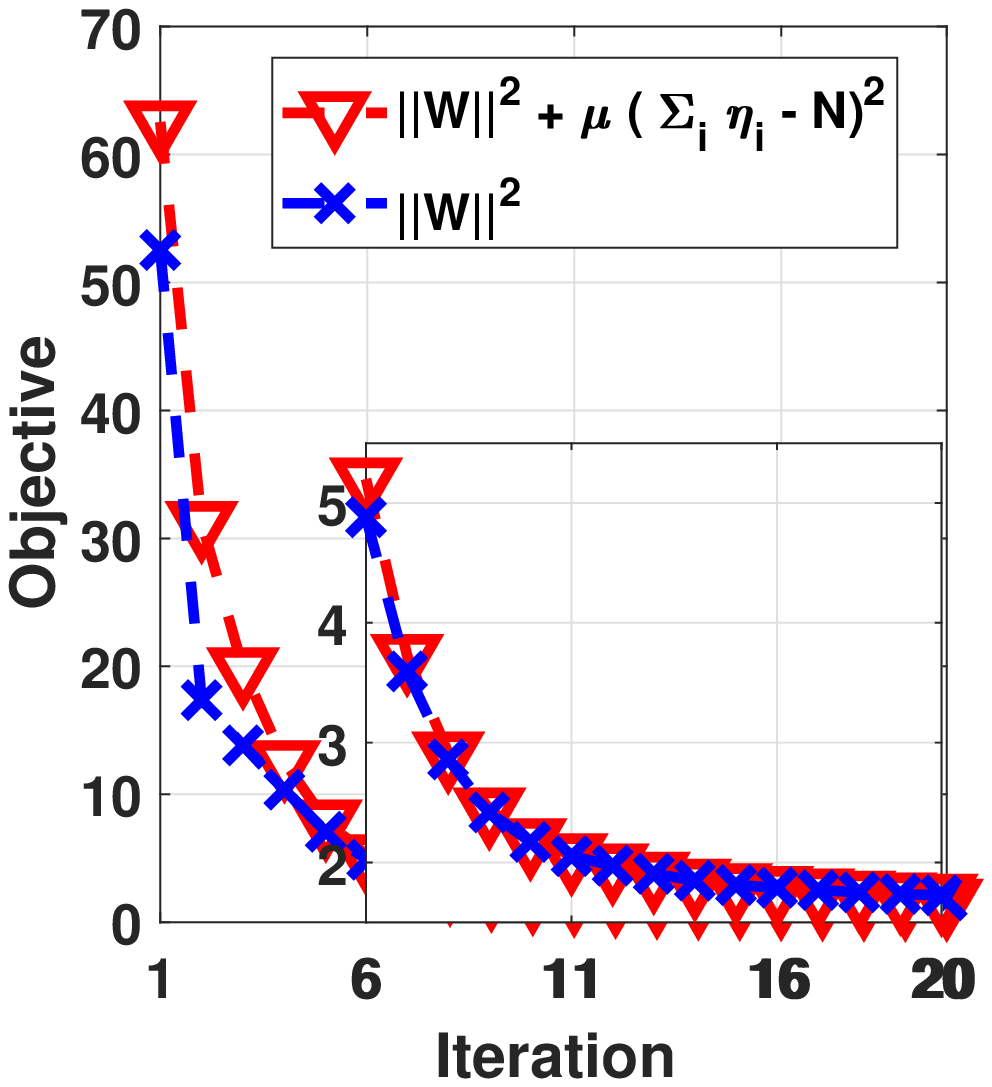}\label{fig:OBJ_PMIN}}
        \caption{Performance comparison of-of PMIN for $\ntxants=10$, $\totpow=10$dB and $\lbrace \epsilon_i=0\text{dB}\rbrace_{i=1}^{\nuserpercell}$.}\label{fig:Objval_PMIN}
    \end{figure}

     The performance JSP-PMIN for uniform weighted case (i.e., all users with same minimum SINR requirement) is illustrated in figure~\ref{fig:Objval_PMIN} for $\ntxants=10$, $\totpow=10$dB and $\lbrace \epsilon_i=1\rbrace_{i=1}^{\nuserpercell}$. In figure~\ref{fig:PMINUnwegihted}, the average total power per CR in dB is depicted as a function of $\nuserpercell$ varying from 15 to 30 in steps of 5. The superior performance of JSP-PMIN over SUS-PMIN is clear from figure~\ref{fig:PMINUnwegihted}. However, the gains diminish as $\nuserpercell$ increases as the SUS based scheduling becomes efficient as mentioned previously. In figure~\ref{fig:OBJ_PMIN}, the convergence behavior of the JSP-PMIN algorithm (red curve) and the progression of ensuring the exact scheduling constraint is depicted as a function of iterations for $\nuserpercell=15$. The FIP may include the solutions that violate exact scheduling constraint due to which the penalized objective and objective differs by a large factor in the initial iterations. However, the increment in the penalty parameter $\mu$ ensures the exact scheduling constraint over the iterations. This is confirmed by figure~\ref{fig:OBJ_PMIN}, as the difference between penalized objective and objective, becomes approximately zero. For the reasons at the beginning of this section, $\lambda_3=0$ still achieves the binary nature of $\boldsymbol{\eta}$ over iterations.

    \section{Conclusions}\label{sec:conclusions}
    In this paper, the joint scheduling and precoding problem was considered for multiuser MISO downlink channels for three different criteria (weighted sum rate maximization, maximization of minimum SINR and power minimization). Unlike the existing works, the design is formulated in a way that is amenable to the joint update scheduling and precoding. Noticing that the original optimization to be MINLP problems in all the cases, we have proposed efficient reformulations and relaxations to transform these into structured DC programming problems. Subsequently, we proposed joint scheduling and precoding algorithms (JSP-WSR, JSP-MMSINR, and JSP-PMIN) for the aforementioned criteria, which are guaranteed to converge to a stationary point. Finally, we propose a simple procedure to obtain a good feasible initial point, critical to the implementation of CCP based algorithms. Through simulations, we established the efficacy of the proposed joint techniques with respect to the decoupled benchmark solutions.

    

\bibliographystyle{IEEEtran}
\bibliography{IEEEabrv,bibJournalList,JSP_Journal_draft}

\begin{thebibliography}{10}
\providecommand{\url}[1]{#1}
\csname url@samestyle\endcsname
\providecommand{\newblock}{\relax}
\providecommand{\bibinfo}[2]{#2}
\providecommand{\BIBentrySTDinterwordspacing}{\spaceskip=0pt\relax}
\providecommand{\BIBentryALTinterwordstretchfactor}{4}
\providecommand{\BIBentryALTinterwordspacing}{\spaceskip=\fontdimen2\font plus
\BIBentryALTinterwordstretchfactor\fontdimen3\font minus
  \fontdimen4\font\relax}
\providecommand{\BIBforeignlanguage}[2]{{%
\expandafter\ifx\csname l@#1\endcsname\relax
\typeout{** WARNING: IEEEtran.bst: No hyphenation pattern has been}%
\typeout{** loaded for the language `#1'. Using the pattern for}%
\typeout{** the default language instead.}%
\else
\language=\csname l@#1\endcsname
\fi
#2}}
\providecommand{\BIBdecl}{\relax}
\BIBdecl

\bibitem{GlobalSIP_ref}
A.~Bandi, M.~R.~B. Shakar, S.~Maleki, C.~Symeon, and B.~Ottersten, ``{A Novel
  Approach to Joint User Selection and Precoding for Multiuser MISO Downlink
  Channels},'' in \emph{2018 Proceedings of GlobalSIP}, November 2018.

\bibitem{1683918}
H.~Weingarten, Y.~Steinberg, and S.~S. Shamai, ``{The Capacity Region of the
  Gaussian Multiple-Input Multiple-Output Broadcast Channel},'' \emph{{IEEE}
  Trans. Inf. Theory}, vol.~52, no.~9, pp. 3936--3964, Sept 2006.

\bibitem{Beamforming}
H.~Viswanathan, S.~Venkatesan, and H.~Huang, ``{Downlink capacity evaluation of
  cellular networks with known-interference cancellation},'' \emph{{IEEE} J.
  Sel. Areas Commun.}, vol.~21, no.~5, pp. 802--811, June 2003.

\bibitem{406634}
P.~Zetterberg and B.~Ottersten, ``{The spectrum efficiency of a base station
  antenna array system for spatially selective transmission},'' \emph{{IEEE}
  Trans. Veh. Technol.}, vol.~44, no.~3, pp. 651--660, Aug 1995.

\bibitem{69993}
S.~Anderson, M.~Millnert, M.~Viberg, and B.~Wahlberg, ``{An adaptive array for
  mobile communication systems},'' \emph{{IEEE} Trans. Veh. Technol.}, vol.~40,
  no.~1, pp. 230--236, Feb 1991.

\bibitem{GreedyUS_Dimic}
G.~Dimic and N.~D. Sidiropoulos, ``{On downlink beamforming with greedy user
  selection: performance analysis and a simple new algorithm},'' \emph{{IEEE}
  Trans. Signal Process.}, vol.~53, no.~10, pp. 3857--3868, Oct 2005.

\bibitem{ChanOrthoZFBF_goldsmith}
T.~Yoo and A.~Goldsmith, ``{On the optimality of multiantenna broadcast
  scheduling using zero-forcing beamforming},'' \emph{{IEEE} J. Sel. Areas
  Commun.}, vol.~24, no.~3, pp. 528--541, March 2006.

\bibitem{ResrcAllcReview_2017}
E.~Castañeda, A.~Silva, A.~Gameiro, and M.~Kountouris, ``{An Overview on
  Resource Allocation Techniques for Multi-User MIMO Systems},'' \emph{IEEE
  Communications Surveys Tutorials}, vol.~19, no.~1, pp. 239--284, Firstquarter
  2017.

\bibitem{SUS_limitedfeedback}
T.~Yoo, N.~Jindal, and A.~Goldsmith, ``{Multi-Antenna Downlink Channels with
  Limited Feedback and User Selection},'' \emph{{IEEE} J. Sel. Areas Commun.},
  vol.~25, no.~7, pp. 1478--1491, September 2007.

\bibitem{8239622}
G.~Lee and Y.~Sung, ``{A New Approach to User Scheduling in Massive Multi-User
  MIMO Broadcast Channels},'' \emph{IEEE Transactions on Communications},
  vol.~66, no.~4, pp. 1481--1495, April 2018.

\bibitem{MMSINR_ref}
B.~Song, Y.~Lin, and R.~L. Cruz, ``{Weighted max-min fair beamforming, power
  control, and scheduling for a MISO downlink},'' \emph{IEEE Transactions on
  Wireless Communications}, vol.~7, no.~2, pp. 464--469, February 2008.

\bibitem{WYU_Multicell_JointSchPrec}
W.~Yu, T.~Kwon, and C.~Shin, ``{Multicell Coordination via Joint Scheduling,
  Beamforming, and Power Spectrum Adaptation},'' \emph{{IEEE} Trans. Wireless
  Commun.}, vol.~12, no.~7, pp. 1--14, July 2013.

\bibitem{MINLP_ref1}
E.~Matskani, N.~D. Sidiropoulos, Z.~q.~Luo, and L.~Tassiulas, ``{Convex
  approximation techniques for joint multiuser downlink beamforming and
  admission control},'' \emph{{IEEE} Trans. Wireless Commun.}, vol.~7, no.~7,
  pp. 2682--2693, July 2008.

\bibitem{MulticellSUS}
M.~Li, I.~B. Collings, S.~V. Hanly, C.~Liu, and P.~Whiting, ``{Multicell
  Coordinated Scheduling With Multiuser Zero-Forcing Beamforming},''
  \emph{{IEEE} Trans. Wireless Commun.}, vol.~15, no.~2, pp. 827--842, Feb
  2016.

\bibitem{4641963}
M.~Kountouris, D.~Gesbert, and T.~Sälzer, ``{Enhanced multiuser random
  beamforming: dealing with the not so large number of users case},''
  \emph{IEEE Journal on Selected Areas in Communications}, vol.~26, no.~8, pp.
  1536--1545, October 2008.

\bibitem{SparsePrecPenalty_JointSchBF_JSAC}
M.~L. Ku, L.~C. Wang, and Y.~L. Liu, ``{Joint Antenna Beamforming, Multiuser
  Scheduling, and Power Allocation for Hierarchical Cellular Systems},''
  \emph{{IEEE} J. Sel. Areas Commun.}, vol.~33, no.~5, pp. 896--909, May 2015.

\bibitem{6334281}
L.~Yu, E.~Karipidis, and E.~G. Larsson, ``Coordinated scheduling and
  beamforming for multicell spectrum sharing networks using branch and bound,''
  in \emph{2012 Proceedings of EUSIPCO}, Aug 2012, pp. 819--823.

\bibitem{CS_PC_CRN_16}
A.~Douik, H.~Dahrouj, T.~Y. Al-Naffouri, and M.~S. Alouini, ``{Coordinated
  Scheduling and Power Control in Cloud-Radio Access Networks},'' \emph{IEEE
  Transactions on Wireless Communications}, vol.~15, no.~4, pp. 2523--2536,
  April 2016.

\bibitem{6920005}
B.~Dai and W.~Yu, ``{Sparse Beamforming and User-Centric Clustering for
  Downlink Cloud Radio Access Network},'' \emph{IEEE Access}, vol.~2, pp.
  1326--1339, 2014.

\bibitem{SparsePrecodingPenalty_MTao}
M.~Tao, E.~Chen, H.~Zhou, and W.~Yu, ``{Content-Centric Sparse Multicast
  Beamforming for Cache-Enabled Cloud RAN},'' \emph{{IEEE} Trans. Wireless
  Commun.}, vol.~15, no.~9, pp. 6118--6131, Sept 2016.

\bibitem{CC_2017}
S.~He, J.~Wang, Y.~Huang, B.~Ottersten, and W.~Hong, ``{Codebook-Based Hybrid
  Precoding for Millimeter Wave Multiuser Systems},'' \emph{IEEE_J_SP},
  vol.~65, no.~20, pp. 5289--5304, Oct 2017.

\bibitem{6155555}
A.~H. Phan, H.~D. Tuan, H.~H. Kha, and H.~H. Nguyen, ``{Beamforming
  Optimization in Multi-User Amplify-and-Forward Wireless Relay Networks},''
  \emph{IEEE Transactions on Wireless Communications}, vol.~11, no.~4, pp.
  1510--1520, April 2012.

\bibitem{6476606}
U.~Rashid, H.~D. Tuan, and H.~H. Nguyen, ``{Relay Beamforming Designs in
  Multi-User Wireless Relay Networks Based on Throughput Maximin
  Optimization},'' \emph{IEEE Transactions on Communications}, vol.~61, no.~5,
  pp. 1739--1749, May 2013.

\bibitem{Fairness_lit}
H.~Shi, R.~V. Prasad, E.~Onur, and I.~G. M.~M. Niemegeers, ``{Fairness in
  Wireless Networks:Issues, Measures and Challenges}, year={2014},''
  \emph{{IEEE} Commun. Surveys Tuts.}, vol.~16, no.~1, pp. 5--24, First.

\bibitem{MGMC_JSP}
D.~Christopoulos, S.~Chatzinotas, and B.~Ottersten, ``Multicast multigroup
  precoding and user scheduling for frame-based satellite communications,''
  \emph{IEEE Transactions on Wireless Communications}, vol.~14, no.~9, pp.
  4695--4707, Sept 2015.

\bibitem{6059438}
I.~Mitliagkas, N.~D. Sidiropoulos, and A.~Swami, ``{Joint Power and Admission
  Control for Ad-Hoc and Cognitive Underlay Networks: Convex Approximation and
  Distributed Implementation},'' \emph{{IEEE} Trans. Wireless Commun.},
  vol.~10, no.~12, pp. 4110--4121, December 2011.

\bibitem{7012104}
Y.~Cheng and M.~Pesavento, ``{Joint Discrete Rate Adaptation and Downlink
  Beamforming Using Mixed Integer Conic Programming},'' \emph{IEEE Transactions
  on Signal Processing}, vol.~63, no.~7, pp. 1750--1764, April 2015.

\bibitem{7581107}
J.~Rubio, A.~Pascual-Iserte, D.~P. Palomar, and A.~Goldsmith, ``{Joint
  Optimization of Power and Data Transfer in Multiuser MIMO Systems},''
  \emph{IEEE Transactions on Signal Processing}, vol.~65, no.~1, pp. 212--227,
  Jan 2017.

\bibitem{5594709}
C.~T.~K. Ng and H.~Huang, ``{Linear Precoding in Cooperative MIMO Cellular
  Networks with Limited Coordination Clusters},'' \emph{IEEE Journal on
  Selected Areas in Communications}, vol.~28, no.~9, pp. 1446--1454, December
  2010.

\bibitem{WSR_CCP_damped}
S.~You, L.~Chen, and Y.~E. Liu, ``{Convex-concave procedure for weighted
  sum-rate maximization in a MIMO interference network},'' in \emph{2014 IEEE
  Global Communications Conference}, Dec 2014, pp. 4060--4065.

\bibitem{SOCP_Amiweisel}
A.~Wiesel, Y.~C. Eldar, and S.~Shamai, ``Linear precoding via conic
  optimization for fixed mimo receivers,'' \emph{IEEE Transactions on Signal
  Processing}, vol.~54, no.~1, pp. 161--176, Jan 2006.

\bibitem{Bengtsson454064}
M.~Bengtsson and B.~Ottersten, ``Optimal and suboptimal transmit beamforming,''
  in \emph{Handbook of Antennas in Wireless Communications}.\hskip 1em plus
  0.5em minus 0.4em\relax CRC Press, 2001, pp. 18--1--18--33, qC 20111107.

\bibitem{Yuille_CCCP_2001}
A.~L. Yuille and A.~Rangarajan, ``{The concave-convex procedure ({CCCP})},'' in
  \emph{NIPS}, 2001.

\bibitem{Complexity_ref}
P.~Gahinet, A.~Nemirovski, A.~J. Laub, and M.~Chilali, \emph{LMI Control
  Toolbox User’s Guide}.\hskip 1em plus 0.5em minus 0.4em\relax USA:
  MathWorks,, 1995.

\bibitem{Sriperumbudur_CCP_2009}
G.~R. Lanckriet and B.~K. Sriperumbudur, ``{On the Convergence of the
  Concave-Convex Procedure},'' in \emph{Advances in Neural Information
  Processing Systems 22}, 2009, pp. 1759--1767.

\bibitem{5762643}
D.~W.~H. Cai, T.~Q.~S. Quek, and C.~W. Tan, ``{A Unified Analysis of Max-Min
  Weighted SINR for MIMO Downlink System},'' \emph{IEEE Transactions on Signal
  Processing}, vol.~59, no.~8, pp. 3850--3862, Aug 2011.

\bibitem{7583660}
L.~Zheng, Y.~.~P. Hong, C.~W. Tan, C.~Hsieh, and C.~Lee, ``{Wireless Max–Min
  Utility Fairness With General Monotonic Constraints by Perron–Frobenius
  Theory},'' \emph{{IEEE} Trans. Inf. Theory}, vol.~62, no.~12, pp. 7283--7298,
  Dec 2016.

\end{thebibliography}


\end{document}